%% file: mdf.tex
\documentclass{emulateapj}
\usepackage{ifthen,amsmath}

\shorttitle{Metallicity Distributions in MW Dwarf Satellites}
\shortauthors{Kirby et al.}
\slugcomment{Accepted to ApJ, 2010 November 7}

\begin{document}
\newcommand{\teff}{$T_{\rm{eff}}$}
\newcommand{\mathteff}{T_{\rm eff}}
\newcommand{\logg}{$\log g$}
\newcommand{\mathlogg}{\log g}
\newcommand{\feh}{[Fe/H]}
\newcommand{\mathfeh}{{\rm [Fe/H]}}
\newcommand{\afe}{[$\alpha$/Fe]}
\newcommand{\mathafe}{{\rm [\alpha/Fe]}}
\newcommand{\xfe}{[$\alpha$/Fe]}
\newcommand{\mathxfe}{{\rm [X/Fe]}}
\newcommand{\ah}{[$\alpha$/H]}
\newcommand{\mathah}{{\rm [\alpha/H]}}
\newcommand{\vt}{$v_t$}
\newcommand{\mathvt}{v_t}
\newcommand{\ngcaf}{NGC~1904 (M79)}
\newcommand{\ngca}{M79}
\newcommand{\ngcb}{NGC 2419}
\newcommand{\ngccf}{NGC~6205 (M13)}
\newcommand{\ngcc}{M13}
\newcommand{\ngcdf}{NGC~6838 (M71)}
\newcommand{\ngcd}{M71}
\newcommand{\ngce}{NGC 7006}
\newcommand{\ngcff}{NGC~7078 (M15)}
\newcommand{\ngcf}{M15}
\newcommand{\ngcg}{NGC 7492}
\newcommand{\bd}{BD+28$^{\circ}$~4211}
\newcommand{\mathaa}{\mathrm{\AA}}
\newcommand\smion[2]{#1$\;$\protect \footnotesize{#2}\small}%
\newcommand\scion[2]{#1$\;$\protect \tiny{#2}\scriptsize}%

\newcommand{\ndup}{167}
\newcommand{\nalphadup}{141}
\newcommand{\fehdupsigma}{0.96}
\newcommand{\fehdupkurtosis}{-0.14}
\newcommand{\alphadupsigma}{0.91}
\newcommand{\alphadupkurtosis}{-0.05}
\newcommand{\verr}{3.0}
\newcommand{\verrerr}{0.2}
\newcommand{\medtefferrrand}{53}
\newcommand{\medtefferrsys}{60}
\newcommand{\medtefferrtot}{85}
\newcommand{\medloggerrrand}{0.05}
\newcommand{\medloggerrsys}{0.02}
\newcommand{\medloggerrtot}{0.06}
\newcommand{\ndsphstars}{2961}
\newcommand{\foryieldsimple}{0.118}
\newcommand{\foryieldsimpleerr}{0.005}
\newcommand{\foryieldpre}{0.096}
\newcommand{\foryieldpreerr}{0.005}
\newcommand{\forfehpre}{-2.12}
\newcommand{\forfehpreerrhi}{0.07}
\newcommand{\forfehpreerrlo}{0.07}
\newcommand{\foryieldinfall}{0.122}
\newcommand{\foryieldinfallerr}{0.004}
\newcommand{\forminfall}{7.4}
\newcommand{\forminfallerrhi}{1.2}
\newcommand{\forminfallerrlo}{1.0}
\newcommand{\leoiyieldsimple}{0.044}
\newcommand{\leoiyieldsimpleerr}{0.002}
\newcommand{\leoiyieldpre}{0.033}
\newcommand{\leoiyieldpreerr}{0.002}
\newcommand{\leoifehpre}{-2.35}
\newcommand{\leoifehpreerrhi}{0.05}
\newcommand{\leoifehpreerrlo}{0.06}
\newcommand{\leoiyieldinfall}{0.045}
\newcommand{\leoiyieldinfallerr}{0.001}
\newcommand{\leoiminfall}{7.2}
\newcommand{\leoiminfallerrhi}{1.0}
\newcommand{\leoiminfallerrlo}{0.9}
\newcommand{\sclyieldsimple}{0.029}
\newcommand{\sclyieldsimpleerr}{0.002}
\newcommand{\sclyieldpre}{0.028}
\newcommand{\sclyieldpreerr}{0.002}
\newcommand{\sclfehpre}{-3.67}
\newcommand{\sclfehpreerrhi}{0.33}
\newcommand{\sclfehpreerrlo}{14.51}
\newcommand{\sclyieldinfall}{0.029}
\newcommand{\sclyieldinfallerr}{0.002}
\newcommand{\sclminfall}{1.3}
\newcommand{\sclminfallerrhi}{0.2}
\newcommand{\sclminfallerrlo}{0.1}
\newcommand{\leoiiyieldsimple}{0.029}
\newcommand{\leoiiyieldsimpleerr}{0.002}
\newcommand{\leoiiyieldpre}{0.025}
\newcommand{\leoiiyieldpreerr}{0.002}
\newcommand{\leoiifehpre}{-2.94}
\newcommand{\leoiifehpreerrhi}{0.11}
\newcommand{\leoiifehpreerrlo}{0.14}
\newcommand{\leoiiyieldinfall}{0.030}
\newcommand{\leoiiyieldinfallerr}{0.002}
\newcommand{\leoiiminfall}{3.1}
\newcommand{\leoiiminfallerrhi}{0.6}
\newcommand{\leoiiminfallerrlo}{0.5}
\newcommand{\sexyieldsimple}{0.016}
\newcommand{\sexyieldsimpleerr}{0.002}
\newcommand{\sexyieldpre}{0.014}
\newcommand{\sexyieldpreerr}{0.002}
\newcommand{\sexfehpre}{-3.20}
\newcommand{\sexfehpreerrhi}{0.17}
\newcommand{\sexfehpreerrlo}{0.24}
\newcommand{\sexyieldinfall}{0.015}
\newcommand{\sexyieldinfallerr}{0.001}
\newcommand{\sexminfall}{3.0}
\newcommand{\sexminfallerrhi}{1.2}
\newcommand{\sexminfallerrlo}{0.8}
\newcommand{\drayieldsimple}{0.016}
\newcommand{\drayieldsimpleerr}{0.001}
\newcommand{\drayieldpre}{0.013}
\newcommand{\drayieldpreerr}{0.001}
\newcommand{\drafehpre}{-3.05}
\newcommand{\drafehpreerrhi}{0.10}
\newcommand{\drafehpreerrlo}{0.11}
\newcommand{\drayieldinfall}{0.015}
\newcommand{\drayieldinfallerr}{0.001}
\newcommand{\draminfall}{4.0}
\newcommand{\draminfallerrhi}{1.1}
\newcommand{\draminfallerrlo}{0.8}
\newcommand{\cvniyieldsimple}{0.018}
\newcommand{\cvniyieldsimpleerr}{0.002}
\newcommand{\cvniyieldpre}{0.016}
\newcommand{\cvniyieldpreerr}{0.002}
\newcommand{\cvnifehpre}{-3.37}
\newcommand{\cvnifehpreerrhi}{0.20}
\newcommand{\cvnifehpreerrlo}{0.37}
\newcommand{\cvniyieldinfall}{0.017}
\newcommand{\cvniyieldinfallerr}{0.002}
\newcommand{\cvniminfall}{1.6}
\newcommand{\cvniminfallerrhi}{0.5}
\newcommand{\cvniminfallerrlo}{0.3}
\newcommand{\umiyieldsimple}{0.011}
\newcommand{\umiyieldsimpleerr}{0.001}
\newcommand{\umiyieldpre}{0.007}
\newcommand{\umiyieldpreerr}{0.001}
\newcommand{\umifehpre}{-2.91}
\newcommand{\umifehpreerrhi}{0.09}
\newcommand{\umifehpreerrlo}{0.10}
\newcommand{\umiyieldinfall}{0.009}
\newcommand{\umiyieldinfallerr}{0.001}
\newcommand{\umiminfall}{9.1}
\newcommand{\umiminfallerrhi}{4.4}
\newcommand{\umiminfallerrlo}{3.0}
\newcommand{\fehbatsigma}{0.17}
\newcommand{\cafebatsigma}{1.38}
\newcommand{\cafebatcorr}{0.00}
\newcommand{\nbat}{57}
\newcommand{\dsphteffdiffmean}{-40}
\newcommand{\dsphteffdiffsigma}{132}
\newcommand{\dsphloggdiffmean}{+0.05}
\newcommand{\dsphloggdiffsigma}{0.41}
\newcommand{\dsphvtdiffmean}{-0.2}
\newcommand{\dsphvtdiffsigma}{0.3}
\newcommand{\dsphfehdiffmean}{-0.04}
\newcommand{\dsphfehdiffsigma}{0.15}
\newcommand{\dsphmgfediffmean}{-0.06}
\newcommand{\dsphmgfediffsigma}{0.18}
\newcommand{\dsphsifediffmean}{-0.16}
\newcommand{\dsphsifediffsigma}{0.28}
\newcommand{\dsphcafediffmean}{-0.03}
\newcommand{\dsphcafediffsigma}{0.23}
\newcommand{\dsphtifediffmean}{-0.05}
\newcommand{\dsphtifediffsigma}{0.24}
\newcommand{\dsphalphafediffmean}{-0.05}
\newcommand{\dsphalphafediffsigma}{0.19}
\newcommand{\lzrkirby}{\langle{\rm [Fe/H]}\rangle = (-2.04 \pm 0.04) + (0.31 \pm 0.04) \log \left(\frac{L_{\rm tot}}{10^5 L_{\sun}}\right)}
\newcommand{\pearsonkirby}{0.95}
\newcommand{\lzrall}{\langle{\rm [Fe/H]}\rangle = (-2.02 \pm 0.04) + (0.31 \pm 0.04) \log \left(\frac{L_{\rm tot}}{10^5 L_{\sun}}\right)}
\newcommand{\pearsonall}{0.93}
\newcommand{\lzsrkirby}{\sigma({\rm [Fe/H]}) = (0.45 \pm 0.03) - (0.06 \pm 0.02) \log \left(\frac{L_{\rm tot}}{10^5 L_{\sun}}\right)}
\newcommand{\pearsonskirby}{-0.57}
\newcommand{\lzzsrkirby}{\log \sigma(Z/Z_{\sun}) = (-1.90 \pm 0.05) + (0.28 \pm 0.05) \log \left(\frac{L_{\rm tot}}{10^5 L_{\sun}}\right)}
\newcommand{\pearsonszkirby}{ 0.85}
\newcommand{\sclnbad}{1}
\newcommand{\sclngood}{411}
\newcommand{\sclndup}{17}
\newcommand{\sclnunique}{394}
\newcommand{\alphasyserr}{0.082}
\newcommand{\mgfesyserr}{0.095}
\newcommand{\sifesyserr}{0.106}
\newcommand{\cafesyserr}{0.118}
\newcommand{\tifesyserr}{0.083}
\newcommand{\scllowv}{85.1}
\newcommand{\sclhighv}{138.0}
\newcommand{\sclmeanv}{111.6}
\newcommand{\sclmeanverr}{0.5}
\newcommand{\sclvdupsigma}{3.0}
\newcommand{\sclsigmav}{8.3}
\newcommand{\sclsigmaverr}{0.5}
\newcommand{\sclmeanvother}{110.4}
\newcommand{\sclmeanverrother}{0.8}
\newcommand{\sclsigmavother}{8.8}
\newcommand{\sclsigmaverrother}{0.6}
\newcommand{\sclmeanvdiff}{+1.2}
\newcommand{\sclsigmavdiff}{-0.7}
\newcommand{\sclmeanvw}{111.3}
\newcommand{\sclmeanverrw}{0.2}
\newcommand{\sclsigmavw}{8.8}
\newcommand{\sclsigmaverrw}{0.2}
\newcommand{\sclmeanvdiffw}{+0.5}
\newcommand{\sclsigmavdiffw}{-0.9}
\newcommand{\sclalphaerfsigma}{2.4}
\newcommand{\sclnnonmember}{18}
\newcommand{\sclnmember}{376}
\newcommand{\sclnvmp}{115}
\newcommand{\sclmedtefferrrand}{94}
\newcommand{\sclmedtefferrsys}{58}
\newcommand{\sclmedtefferrtot}{110}
\newcommand{\sclmedloggerrrand}{0.06}
\newcommand{\sclmedloggerrsys}{0.02}
\newcommand{\sclmedloggerrtot}{0.06}
\newcommand{\sclfehhrssigma}{0.16}
\newcommand{\sclfehbatsigma}{0.17}
\newcommand{\sclcafebatsigma}{1.47}
\newcommand{\sclcafebatcorr}{****}
\newcommand{\sclnbat}{50}
\newcommand{\sclmdfhelmean}{-1.82}
\newcommand{\sclmdfhelsigma}{0.35}
\newcommand{\sclmdfbatmean}{-1.56}
\newcommand{\sclmdfbatsigma}{0.38}
\newcommand{\sclnbathrs}{7}
\newcommand{\sclfehbathrssigma}{0.16}
\newcommand{\sclfehcuthrssigma}{0.19}
\newcommand{\sclfehslope}{-1.856}
\newcommand{\sclfehslopeerr}{0.160}
\newcommand{\sclfehslopekpc}{-1.24}
\newcommand{\sclfehslopekpcerr}{0.11}
\newcommand{\sclfehsloperc}{-0.18}
\newcommand{\sclfehslopercerr}{0.02}
\newcommand{\sclafeslope}{+0.013}
\newcommand{\sclafeslopeerr}{0.003}
\newcommand{\sclafeslopekpc}{+0.54}
\newcommand{\sclafeslopekpcerr}{0.10}
\newcommand{\sclalphafehrssigma}{0.17}
\newcommand{\sclfehrange}{****}
\newcommand{\sclfehinitial}{-7.81}
\newcommand{\sclfehinitialerr}{0.00}
\newcommand{\sclfehfinal}{-7.82}
\newcommand{\sclfehfinalerr}{0.00}
\newcommand{\sclpureyield}{0.030}
\newcommand{\sclsimpleyield}{0.028}
\newcommand{\sclsimpleyielderr}{0.001}
\newcommand{\sclinfallm}{1.3}
\newcommand{\sclinfallyield}{0.030}
\newcommand{\sclprobratio}{1.59}
\newcommand{\sclprobratiopure}{-0.11}
\newcommand{\sclprobratioinfallpre}{1.70}
\newcommand{\sclprobratiopreinfall}{-1.70}
\newcommand{\sclfehspread}{0.46}
\newcommand{\sclmgfespread}{0.21}
\newcommand{\sclsifespread}{0.29}
\newcommand{\sclcafespread}{0.17}
\newcommand{\scltifespread}{0.22}
\newcommand{\sclalphafespread}{0.24}
\newcommand{\sclmgfespreadsub}{-0.10}
\newcommand{\sclsifespreadsub}{0.10}
\newcommand{\sclcafespreadsub}{0.05}
\newcommand{\scltifespreadsub}{0.09}
\newcommand{\sclalphafespreadsub}{0.09}
\newcommand{\sclfehmean}{-1.68}
\newcommand{\sclfehsigma}{0.48}
\newcommand{\sclfehmedian}{-1.67}
\newcommand{\sclfehmad}{0.37}
\newcommand{\sclfehiqr}{0.75}
\newcommand{\sclempfehone}{-3.02}
\newcommand{\sclempfeherrone}{0.15}
\newcommand{\sclempvone}{18.06}
\newcommand{\sclempfehtwo}{-3.27}
\newcommand{\sclempfeherrtwo}{0.59}
\newcommand{\sclempvtwo}{20.08}
\newcommand{\sclempfehthree}{-3.87}
\newcommand{\sclempfeherrthree}{0.21}
\newcommand{\sclempvthree}{18.19}
\newcommand{\sclcmdfeh}{-1.53}
\newcommand{\fornbad}{17}
\newcommand{\forngood}{701}
\newcommand{\forndup}{15}
\newcommand{\fornunique}{686}
\newcommand{\forlowv}{19.9}
\newcommand{\forhighv}{88.0}
\newcommand{\formeanv}{53.9}
\newcommand{\formeanverr}{0.5}
\newcommand{\forvdupsigma}{5.8}
\newcommand{\forsigmav}{9.7}
\newcommand{\forsigmaverr}{1.5}
\newcommand{\formeanvw}{55.2}
\newcommand{\formeanverrw}{0.2}
\newcommand{\forsigmavw}{12.0}
\newcommand{\forsigmaverrw}{0.2}
\newcommand{\formeanvdiffw}{-2.3}
\newcommand{\forsigmavdiffw}{-1.5}
\newcommand{\foralphaerfsigma}{2.0}
\newcommand{\fornnonmember}{11}
\newcommand{\fornmember}{675}
\newcommand{\fornvmp}{20}
\newcommand{\formedtefferrrand}{49}
\newcommand{\formedtefferrsys}{58}
\newcommand{\formedtefferrtot}{76}
\newcommand{\formedloggerrrand}{0.05}
\newcommand{\formedloggerrsys}{0.03}
\newcommand{\formedloggerrtot}{0.06}
\newcommand{\forfehhrssigma}{0.13}
\newcommand{\forfehbatsigma}{0.10}
\newcommand{\forcafebatsigma}{0.18}
\newcommand{\forcafebatcorr}{****}
\newcommand{\fornbat}{7}
\newcommand{\formdfhelmean}{-1.28}
\newcommand{\formdfhelsigma}{0.45}
\newcommand{\formdfbatmean}{-0.89}
\newcommand{\formdfbatsigma}{0.36}
\newcommand{\fornbathrs}{6}
\newcommand{\forfehbathrssigma}{0.05}
\newcommand{\forfehcuthrssigma}{0.12}
\newcommand{\forfehslope}{-0.073}
\newcommand{\forfehslopeerr}{0.094}
\newcommand{\forfehslopekpc}{-0.03}
\newcommand{\forfehslopekpcerr}{0.04}
\newcommand{\forfehsloperc}{-0.02}
\newcommand{\forfehslopercerr}{0.02}
\newcommand{\forafeslope}{+0.004}
\newcommand{\forafeslopeerr}{0.002}
\newcommand{\forafeslopekpc}{+0.09}
\newcommand{\forafeslopekpcerr}{0.05}
\newcommand{\foralphafehrssigma}{0.16}
\newcommand{\forfehrange}{0.00}
\newcommand{\forfehinitial}{-2.12}
\newcommand{\forfehinitialerr}{0.00}
\newcommand{\forfehfinal}{-2.12}
\newcommand{\forfehfinalerr}{0.00}
\newcommand{\forpureyield}{0.118}
\newcommand{\forsimpleyield}{0.096}
\newcommand{\forsimpleyielderr}{0.001}
\newcommand{\forinfallm}{7.5}
\newcommand{\forinfallyield}{0.122}
\newcommand{\forprobratio}{125.62}
\newcommand{\forprobratiopure}{49.98}
\newcommand{\forprobratioinfallpre}{75.64}
\newcommand{\forprobratiopreinfall}{-75.64}
\newcommand{\forfehspread}{0.31}
\newcommand{\formgfespread}{0.12}
\newcommand{\forsifespread}{0.23}
\newcommand{\forcafespread}{0.23}
\newcommand{\fortifespread}{0.19}
\newcommand{\foralphafespread}{0.22}
\newcommand{\formgfespreadsub}{0.10}
\newcommand{\forsifespreadsub}{0.20}
\newcommand{\forcafespreadsub}{0.22}
\newcommand{\fortifespreadsub}{0.17}
\newcommand{\foralphafespreadsub}{0.19}
\newcommand{\forfehmean}{-0.99}
\newcommand{\forfehsigma}{0.36}
\newcommand{\forfehmedian}{-1.01}
\newcommand{\forfehmad}{0.19}
\newcommand{\forfehiqr}{0.37}
\newcommand{\forcmdfeh}{-1.53}
\newcommand{\leoinbad}{6}
\newcommand{\leoingood}{910}
\newcommand{\leoindup}{44}
\newcommand{\leoinunique}{866}
\newcommand{\leoilowv}{257.4}
\newcommand{\leoihighv}{315.3}
\newcommand{\leoimeanv}{286.3}
\newcommand{\leoimeanverr}{0.3}
\newcommand{\leoivdupsigma}{2.4}
\newcommand{\leoisigmav}{9.3}
\newcommand{\leoisigmaverr}{0.3}
\newcommand{\leoialphaerfsigma}{1.8}
\newcommand{\leoinnonmember}{39}
\newcommand{\leoinmember}{827}
\newcommand{\leoinvmp}{43}
\newcommand{\leoimedtefferrrand}{56}
\newcommand{\leoimedtefferrsys}{52}
\newcommand{\leoimedtefferrtot}{78}
\newcommand{\leoimedloggerrrand}{0.05}
\newcommand{\leoimedloggerrsys}{0.01}
\newcommand{\leoimedloggerrtot}{0.06}
\newcommand{\leoifehslope}{-2.011}
\newcommand{\leoifehslopeerr}{0.101}
\newcommand{\leoifehslopekpc}{-0.45}
\newcommand{\leoifehslopekpcerr}{0.02}
\newcommand{\leoifehsloperc}{-0.11}
\newcommand{\leoifehslopercerr}{0.01}
\newcommand{\leoiafeslope}{+0.009}
\newcommand{\leoiafeslopeerr}{0.002}
\newcommand{\leoiafeslopekpc}{+0.12}
\newcommand{\leoiafeslopekpcerr}{0.03}
\newcommand{\leoifehrange}{0.00}
\newcommand{\leoifehinitial}{-2.35}
\newcommand{\leoifehinitialerr}{0.00}
\newcommand{\leoifehfinal}{-2.35}
\newcommand{\leoifehfinalerr}{0.00}
\newcommand{\leoipureyield}{0.044}
\newcommand{\leoisimpleyield}{0.033}
\newcommand{\leoisimpleyielderr}{0.001}
\newcommand{\leoiinfallm}{7.3}
\newcommand{\leoiinfallyield}{0.045}
\newcommand{\leoiprobratio}{163.32}
\newcommand{\leoiprobratiopure}{81.45}
\newcommand{\leoiprobratioinfallpre}{81.87}
\newcommand{\leoiprobratiopreinfall}{-81.87}
\newcommand{\leoifehspread}{0.29}
\newcommand{\leoimgfespread}{0.10}
\newcommand{\leoisifespread}{0.21}
\newcommand{\leoicafespread}{-0.04}
\newcommand{\leoitifespread}{0.13}
\newcommand{\leoialphafespread}{0.15}
\newcommand{\leoimgfespreadsub}{-0.07}
\newcommand{\leoisifespreadsub}{0.13}
\newcommand{\leoicafespreadsub}{-0.07}
\newcommand{\leoitifespreadsub}{0.12}
\newcommand{\leoialphafespreadsub}{0.11}
\newcommand{\leoifehmean}{-1.43}
\newcommand{\leoifehsigma}{0.33}
\newcommand{\leoifehmedian}{-1.42}
\newcommand{\leoifehmad}{0.18}
\newcommand{\leoifehiqr}{0.37}
\newcommand{\leoiempfehone}{-3.01}
\newcommand{\leoiempfeherrone}{0.29}
\newcommand{\leoiempvone}{21.26}
\newcommand{\leoiempfehtwo}{-3.14}
\newcommand{\leoiempfeherrtwo}{0.52}
\newcommand{\leoiempvtwo}{21.32}
\newcommand{\leoiempfehthree}{-3.22}
\newcommand{\leoiempfeherrthree}{0.78}
\newcommand{\leoiempvthree}{21.35}
\newcommand{\leoiempfehfour}{-3.30}
\newcommand{\leoiempfeherrfour}{0.18}
\newcommand{\leoiempvfour}{20.25}
\newcommand{\leoicmdfeh}{-1.53}
\newcommand{\sexnbad}{49}
\newcommand{\sexngood}{209}
\newcommand{\sexndup}{4}
\newcommand{\sexnunique}{205}
\newcommand{\sexlowv}{199.8}
\newcommand{\sexhighv}{250.5}
\newcommand{\sexmeanv}{225.1}
\newcommand{\sexmeanverr}{0.7}
\newcommand{\sexvdupsigma}{1.6}
\newcommand{\sexsigmav}{8.3}
\newcommand{\sexsigmaverr}{0.6}
\newcommand{\sexmeanvw}{224.2}
\newcommand{\sexmeanverrw}{0.4}
\newcommand{\sexsigmavw}{8.0}
\newcommand{\sexsigmaverrw}{0.5}
\newcommand{\sexmeanvdiffw}{+1.1}
\newcommand{\sexsigmavdiffw}{+0.3}
\newcommand{\sexalphaerfsigma}{0.4}
\newcommand{\sexnnonmember}{64}
\newcommand{\sexnmember}{141}
\newcommand{\sexnvmp}{71}
\newcommand{\sexmedtefferrrand}{50}
\newcommand{\sexmedtefferrsys}{77}
\newcommand{\sexmedtefferrtot}{95}
\newcommand{\sexmedloggerrrand}{0.05}
\newcommand{\sexmedloggerrsys}{0.02}
\newcommand{\sexmedloggerrtot}{0.05}
\newcommand{\sexfehhrssigma}{0.13}
\newcommand{\sexfehslope}{+0.202}
\newcommand{\sexfehslopeerr}{0.189}
\newcommand{\sexfehslopekpc}{+0.12}
\newcommand{\sexfehslopekpcerr}{0.11}
\newcommand{\sexfehsloperc}{+0.06}
\newcommand{\sexfehslopercerr}{0.05}
\newcommand{\sexafeslope}{-0.006}
\newcommand{\sexafeslopeerr}{0.004}
\newcommand{\sexafeslopekpc}{-0.22}
\newcommand{\sexafeslopekpcerr}{0.14}
\newcommand{\sexalphafehrssigma}{0.11}
\newcommand{\sexfehrange}{0.02}
\newcommand{\sexfehinitial}{-3.24}
\newcommand{\sexfehinitialerr}{0.00}
\newcommand{\sexfehfinal}{-3.22}
\newcommand{\sexfehfinalerr}{0.00}
\newcommand{\sexpureyield}{0.016}
\newcommand{\sexsimpleyield}{0.014}
\newcommand{\sexsimpleyielderr}{0.001}
\newcommand{\sexinfallm}{3.2}
\newcommand{\sexinfallyield}{0.015}
\newcommand{\sexprobratio}{5.62}
\newcommand{\sexprobratiopure}{6.43}
\newcommand{\sexprobratioinfallpre}{-0.81}
\newcommand{\sexprobratiopreinfall}{0.81}
\newcommand{\sexfehspread}{0.39}
\newcommand{\sexmgfespread}{0.26}
\newcommand{\sexsifespread}{0.38}
\newcommand{\sexcafespread}{-0.04}
\newcommand{\sextifespread}{0.24}
\newcommand{\sexalphafespread}{0.34}
\newcommand{\sexmgfespreadsub}{-0.13}
\newcommand{\sexsifespreadsub}{0.28}
\newcommand{\sexcafespreadsub}{-0.09}
\newcommand{\sextifespreadsub}{0.18}
\newcommand{\sexalphafespreadsub}{0.28}
\newcommand{\sexfehmean}{-1.93}
\newcommand{\sexfehsigma}{0.48}
\newcommand{\sexfehmedian}{-2.00}
\newcommand{\sexfehmad}{0.29}
\newcommand{\sexfehiqr}{0.57}
\newcommand{\sexempfehone}{-3.01}
\newcommand{\sexempfeherrone}{0.28}
\newcommand{\sexempvone}{19.79}
\newcommand{\sexempfehtwo}{-3.03}
\newcommand{\sexempfeherrtwo}{0.20}
\newcommand{\sexempvtwo}{19.11}
\newcommand{\sexempfehthree}{-3.30}
\newcommand{\sexempfeherrthree}{0.58}
\newcommand{\sexempvthree}{20.45}
\newcommand{\sexempfehfour}{-3.40}
\newcommand{\sexempfeherrfour}{0.48}
\newcommand{\sexempvfour}{20.57}
\newcommand{\sexcmdfeh}{-1.53}
\newcommand{\leoiinbad}{10}
\newcommand{\leoiingood}{319}
\newcommand{\leoiindup}{35}
\newcommand{\leoiinunique}{284}
\newcommand{\leoiilowv}{54.8}
\newcommand{\leoiihighv}{102.1}
\newcommand{\leoiimeanv}{78.4}
\newcommand{\leoiimeanverr}{0.5}
\newcommand{\leoiivdupsigma}{1.5}
\newcommand{\leoiisigmav}{7.7}
\newcommand{\leoiisigmaverr}{0.4}
\newcommand{\leoiialphaerfsigma}{2.4}
\newcommand{\leoiinnonmember}{26}
\newcommand{\leoiinmember}{258}
\newcommand{\leoiinvmp}{50}
\newcommand{\leoiimedtefferrrand}{83}
\newcommand{\leoiimedtefferrsys}{62}
\newcommand{\leoiimedtefferrtot}{106}
\newcommand{\leoiimedloggerrrand}{0.05}
\newcommand{\leoiimedloggerrsys}{0.01}
\newcommand{\leoiimedloggerrtot}{0.05}
\newcommand{\leoiifehslope}{-4.257}
\newcommand{\leoiifehslopeerr}{0.306}
\newcommand{\leoiifehslopekpc}{-1.11}
\newcommand{\leoiifehslopekpcerr}{0.08}
\newcommand{\leoiifehsloperc}{-0.21}
\newcommand{\leoiifehslopercerr}{0.01}
\newcommand{\leoiiafeslope}{+0.025}
\newcommand{\leoiiafeslopeerr}{0.007}
\newcommand{\leoiiafeslopekpc}{+0.39}
\newcommand{\leoiiafeslopekpcerr}{0.10}
\newcommand{\leoiifehrange}{0.00}
\newcommand{\leoiifehinitial}{-2.96}
\newcommand{\leoiifehinitialerr}{0.00}
\newcommand{\leoiifehfinal}{-2.96}
\newcommand{\leoiifehfinalerr}{0.00}
\newcommand{\leoiipureyield}{0.029}
\newcommand{\leoiisimpleyield}{0.025}
\newcommand{\leoiisimpleyielderr}{0.001}
\newcommand{\leoiiinfallm}{3.1}
\newcommand{\leoiiinfallyield}{0.030}
\newcommand{\leoiiprobratio}{21.08}
\newcommand{\leoiiprobratiopure}{12.67}
\newcommand{\leoiiprobratioinfallpre}{8.41}
\newcommand{\leoiiprobratiopreinfall}{-8.41}
\newcommand{\leoiifehspread}{0.37}
\newcommand{\leoiimgfespread}{0.23}
\newcommand{\leoiisifespread}{0.24}
\newcommand{\leoiicafespread}{-0.04}
\newcommand{\leoiitifespread}{0.14}
\newcommand{\leoiialphafespread}{0.20}
\newcommand{\leoiimgfespreadsub}{0.06}
\newcommand{\leoiisifespreadsub}{0.14}
\newcommand{\leoiicafespreadsub}{-0.08}
\newcommand{\leoiitifespreadsub}{0.10}
\newcommand{\leoiialphafespreadsub}{0.13}
\newcommand{\leoiifehmean}{-1.62}
\newcommand{\leoiifehsigma}{0.42}
\newcommand{\leoiifehmedian}{-1.59}
\newcommand{\leoiifehmad}{0.23}
\newcommand{\leoiifehiqr}{0.51}
\newcommand{\leoiiempfehone}{-3.09}
\newcommand{\leoiiempfeherrone}{0.22}
\newcommand{\leoiiempvone}{19.82}
\newcommand{\leoiiempfehtwo}{-3.11}
\newcommand{\leoiiempfeherrtwo}{0.33}
\newcommand{\leoiiempvtwo}{20.68}
\newcommand{\leoiiempfehthree}{-3.22}
\newcommand{\leoiiempfeherrthree}{0.34}
\newcommand{\leoiiempvthree}{20.46}
\newcommand{\leoiicmdfeh}{-1.53}
\newcommand{\uminbad}{63}
\newcommand{\umingood}{269}
\newcommand{\umindup}{15}
\newcommand{\uminunique}{254}
\newcommand{\umilowv}{-268.9}
\newcommand{\umihighv}{-216.6}
\newcommand{\umimeanv}{-242.7}
\newcommand{\umimeanverr}{0.6}
\newcommand{\umivdupsigma}{2.3}
\newcommand{\umisigmav}{8.4}
\newcommand{\umisigmaverr}{0.6}
\newcommand{\umimeanvother}{-247.4}
\newcommand{\umimeanverrother}{1.0}
\newcommand{\umisigmavother}{8.8}
\newcommand{\umisigmaverrother}{0.8}
\newcommand{\umimeanvdiff}{+4.0}
\newcommand{\umisigmavdiff}{-0.4}
\newcommand{\umialphaerfsigma}{1.4}
\newcommand{\uminnonmember}{42}
\newcommand{\uminmember}{212}
\newcommand{\uminvmp}{136}
\newcommand{\umimedtefferrrand}{51}
\newcommand{\umimedtefferrsys}{64}
\newcommand{\umimedtefferrtot}{85}
\newcommand{\umimedloggerrrand}{0.05}
\newcommand{\umimedloggerrsys}{0.02}
\newcommand{\umimedloggerrtot}{0.05}
\newcommand{\umifehhrssigma}{0.15}
\newcommand{\umifehslope}{-0.214}
\newcommand{\umifehslopeerr}{0.191}
\newcommand{\umifehslopekpc}{-0.18}
\newcommand{\umifehslopekpcerr}{0.16}
\newcommand{\umifehsloperc}{-0.06}
\newcommand{\umifehslopercerr}{0.05}
\newcommand{\umiafeslope}{+0.001}
\newcommand{\umiafeslopeerr}{0.004}
\newcommand{\umiafeslopekpc}{+0.04}
\newcommand{\umiafeslopekpcerr}{0.21}
\newcommand{\umialphafehrssigma}{0.31}
\newcommand{\umifehrange}{0.00}
\newcommand{\umifehinitial}{-2.91}
\newcommand{\umifehinitialerr}{0.00}
\newcommand{\umifehfinal}{-2.91}
\newcommand{\umifehfinalerr}{0.00}
\newcommand{\umipureyield}{0.011}
\newcommand{\umisimpleyield}{0.007}
\newcommand{\umisimpleyielderr}{0.001}
\newcommand{\umiinfallm}{9.7}
\newcommand{\umiinfallyield}{0.009}
\newcommand{\umiprobratio}{21.63}
\newcommand{\umiprobratiopure}{22.10}
\newcommand{\umiprobratioinfallpre}{-0.47}
\newcommand{\umiprobratiopreinfall}{0.47}
\newcommand{\umifehspread}{0.34}
\newcommand{\umimgfespread}{0.18}
\newcommand{\umisifespread}{0.18}
\newcommand{\umicafespread}{0.11}
\newcommand{\umitifespread}{0.24}
\newcommand{\umialphafespread}{0.22}
\newcommand{\umimgfespreadsub}{-0.09}
\newcommand{\umisifespreadsub}{0.13}
\newcommand{\umicafespreadsub}{0.03}
\newcommand{\umitifespreadsub}{0.17}
\newcommand{\umialphafespreadsub}{0.18}
\newcommand{\umifehmean}{-2.13}
\newcommand{\umifehsigma}{0.47}
\newcommand{\umifehmedian}{-2.13}
\newcommand{\umifehmad}{0.25}
\newcommand{\umifehiqr}{0.50}
\newcommand{\umiempfehone}{-3.03}
\newcommand{\umiempfeherrone}{0.68}
\newcommand{\umiempvone}{21.21}
\newcommand{\umiempfehtwo}{-3.08}
\newcommand{\umiempfeherrtwo}{0.59}
\newcommand{\umiempvtwo}{20.48}
\newcommand{\umiempfehthree}{-3.10}
\newcommand{\umiempfeherrthree}{0.67}
\newcommand{\umiempvthree}{21.29}
\newcommand{\umiempfehfour}{-3.18}
\newcommand{\umiempfeherrfour}{0.85}
\newcommand{\umiempvfour}{21.12}
\newcommand{\umicmdfeh}{-1.53}
\newcommand{\dranbad}{71}
\newcommand{\drangood}{431}
\newcommand{\drandup}{7}
\newcommand{\dranunique}{424}
\newcommand{\dralowv}{-321.0}
\newcommand{\drahighv}{-263.2}
\newcommand{\drameanv}{-292.1}
\newcommand{\drameanverr}{0.6}
\newcommand{\dravdupsigma}{5.3}
\newcommand{\drasigmav}{8.1}
\newcommand{\drasigmaverr}{2.4}
\newcommand{\draalphaerfsigma}{0.7}
\newcommand{\drannonmember}{126}
\newcommand{\dranmember}{298}
\newcommand{\dranvmp}{130}
\newcommand{\dramedtefferrrand}{50}
\newcommand{\dramedtefferrsys}{75}
\newcommand{\dramedtefferrtot}{91}
\newcommand{\dramedloggerrrand}{0.06}
\newcommand{\dramedloggerrsys}{0.02}
\newcommand{\dramedloggerrtot}{0.06}
\newcommand{\drafehhrssigma}{0.09}
\newcommand{\drafehslope}{-0.727}
\newcommand{\drafehslopeerr}{0.131}
\newcommand{\drafehslopekpc}{-0.45}
\newcommand{\drafehslopekpcerr}{0.08}
\newcommand{\drafehsloperc}{-0.11}
\newcommand{\drafehslopercerr}{0.02}
\newcommand{\draafeslope}{+0.010}
\newcommand{\draafeslopeerr}{0.001}
\newcommand{\draafeslopekpc}{+0.36}
\newcommand{\draafeslopekpcerr}{0.05}
\newcommand{\draalphafehrssigma}{0.12}
\newcommand{\drafehrange}{0.00}
\newcommand{\drafehinitial}{-3.06}
\newcommand{\drafehinitialerr}{0.00}
\newcommand{\drafehfinal}{-3.06}
\newcommand{\drafehfinalerr}{0.00}
\newcommand{\drapureyield}{0.016}
\newcommand{\drasimpleyield}{0.013}
\newcommand{\drasimpleyielderr}{0.001}
\newcommand{\drainfallm}{4.1}
\newcommand{\drainfallyield}{0.015}
\newcommand{\draprobratio}{20.34}
\newcommand{\draprobratiopure}{17.51}
\newcommand{\draprobratioinfallpre}{2.84}
\newcommand{\draprobratiopreinfall}{-2.84}
\newcommand{\drafehspread}{0.36}
\newcommand{\dramgfespread}{0.20}
\newcommand{\drasifespread}{0.24}
\newcommand{\dracafespread}{0.11}
\newcommand{\dratifespread}{0.21}
\newcommand{\draalphafespread}{0.26}
\newcommand{\dramgfespreadsub}{0.18}
\newcommand{\drasifespreadsub}{0.12}
\newcommand{\dracafespreadsub}{0.08}
\newcommand{\dratifespreadsub}{0.18}
\newcommand{\draalphafespreadsub}{0.21}
\newcommand{\drafehmean}{-1.93}
\newcommand{\drafehsigma}{0.47}
\newcommand{\drafehmedian}{-1.93}
\newcommand{\drafehmad}{0.26}
\newcommand{\drafehiqr}{0.51}
\newcommand{\draempfehone}{-3.00}
\newcommand{\draempfeherrone}{0.16}
\newcommand{\draempvone}{18.67}
\newcommand{\draempfehtwo}{-3.02}
\newcommand{\draempfeherrtwo}{0.25}
\newcommand{\draempvtwo}{19.22}
\newcommand{\draempfehthree}{-3.05}
\newcommand{\draempfeherrthree}{0.43}
\newcommand{\draempvthree}{20.37}
\newcommand{\draempfehfour}{-3.08}
\newcommand{\draempfeherrfour}{0.98}
\newcommand{\draempvfour}{21.06}
\newcommand{\dracmdfeh}{-1.53}
\newcommand{\cvninbad}{82}
\newcommand{\cvningood}{390}
\newcommand{\cvnindup}{216}
\newcommand{\cvninunique}{174}
\newcommand{\cvnilowv}{4.2}
\newcommand{\cvnihighv}{56.3}
\newcommand{\cvnimeanv}{30.2}
\newcommand{\cvnimeanverr}{0.7}
\newcommand{\cvnivdupsigma}{0.7}
\newcommand{\cvnisigmav}{8.7}
\newcommand{\cvnisigmaverr}{0.6}
\newcommand{\cvnialphaerfsigma}{2.2}
\newcommand{\cvninnonmember}{0}
\newcommand{\cvninmember}{174}
\newcommand{\cvninvmp}{84}
\newcommand{\cvnimedtefferrrand}{231}
\newcommand{\cvnimedtefferrsys}{71}
\newcommand{\cvnimedtefferrtot}{243}
\newcommand{\cvnimedloggerrrand}{0.07}
\newcommand{\cvnimedloggerrsys}{0.02}
\newcommand{\cvnimedloggerrtot}{0.07}
\newcommand{\cvnifehslope}{-0.480}
\newcommand{\cvnifehslopeerr}{0.361}
\newcommand{\cvnifehslopekpc}{-0.13}
\newcommand{\cvnifehslopekpcerr}{0.10}
\newcommand{\cvnifehsloperc}{-0.07}
\newcommand{\cvnifehslopercerr}{0.05}
\newcommand{\cvniafeslope}{+0.011}
\newcommand{\cvniafeslopeerr}{0.009}
\newcommand{\cvniafeslopekpc}{+0.18}
\newcommand{\cvniafeslopekpcerr}{0.14}
\newcommand{\cvnifehrange}{0.00}
\newcommand{\cvnifehinitial}{-3.52}
\newcommand{\cvnifehinitialerr}{0.00}
\newcommand{\cvnifehfinal}{-3.52}
\newcommand{\cvnifehfinalerr}{0.00}
\newcommand{\cvnipureyield}{0.018}
\newcommand{\cvnisimpleyield}{0.016}
\newcommand{\cvnisimpleyielderr}{0.001}
\newcommand{\cvniinfallm}{2.1}
\newcommand{\cvniinfallyield}{0.017}
\newcommand{\cvniprobratio}{0.26}
\newcommand{\cvniprobratiopure}{4.55}
\newcommand{\cvniprobratioinfallpre}{-4.28}
\newcommand{\cvniprobratiopreinfall}{4.28}
\newcommand{\cvnifehspread}{0.44}
\newcommand{\cvnimgfespread}{0.20}
\newcommand{\cvnisifespread}{0.22}
\newcommand{\cvnicafespread}{0.17}
\newcommand{\cvnitifespread}{0.24}
\newcommand{\cvnialphafespread}{0.22}
\newcommand{\cvnimgfespreadsub}{0.05}
\newcommand{\cvnisifespreadsub}{0.13}
\newcommand{\cvnicafespreadsub}{0.09}
\newcommand{\cvnitifespreadsub}{0.19}
\newcommand{\cvnialphafespreadsub}{0.19}
\newcommand{\cvnifehmean}{-1.98}
\newcommand{\cvnifehsigma}{0.55}
\newcommand{\cvnifehmedian}{-1.98}
\newcommand{\cvnifehmad}{0.36}
\newcommand{\cvnifehiqr}{0.71}
\newcommand{\cvniempfehone}{-3.05}
\newcommand{\cvniempfeherrone}{0.56}
\newcommand{\cvniempvone}{22.05}
\newcommand{\cvniempfehtwo}{-3.12}
\newcommand{\cvniempfeherrtwo}{0.81}
\newcommand{\cvniempvtwo}{21.97}
\newcommand{\cvniempfehthree}{-3.13}
\newcommand{\cvniempfeherrthree}{0.66}
\newcommand{\cvniempvthree}{21.66}
\newcommand{\cvniempfehfour}{-3.19}
\newcommand{\cvniempfeherrfour}{0.89}
\newcommand{\cvniempvfour}{21.88}
\newcommand{\cvnicmdfeh}{-1.53}

\title{Multi-Element Abundance Measurements from Medium-Resolution
Spectra. \\ III. Metallicity Distributions of Milky Way Dwarf
Satellite Galaxies}

\author{Evan~N.~Kirby\altaffilmark{2,3},
  Gustavo~A.~Lanfranchi\altaffilmark{4},
  Joshua~D.~Simon\altaffilmark{5},
  Judith~G.~Cohen\altaffilmark{2},
  Puragra~Guhathakurta\altaffilmark{6}}

\altaffiltext{1}{Data herein were obtained at the W.~M. Keck
  Observatory, which is operated as a scientific partnership among the
  California Institute of Technology, the University of California,
  and NASA.  The Observatory was made possible by the generous
  financial support of the W.~M. Keck Foundation.}
\altaffiltext{2}{California Institute of Technology, Department of
  Astronomy, Mail Stop 249-17, Pasadena, CA 91125}
\altaffiltext{3}{Hubble Fellow}
\altaffiltext{4}{N{\' u}cleo de Astrof{\' i}sica Te{\' o}rica,
  Universidade Cruzeiro do Sul, R.~Galv{\~ a}o Bueno 868, Liberdade,
  01506-000, S{\~ a}o Paulo, SP, Brazil}
\altaffiltext{5}{Observatories of the Carnegie Institution of
  Washington, 813 Santa Barbara Street, Pasadena, CA 91101}
\altaffiltext{6}{University of California Observatories/Lick
  Observatory, Department of Astronomy \& Astrophysics, University of
  California, Santa Cruz, CA 95064}



\begin{abstract}
We present metallicity distribution functions (MDFs) for the central
regions of eight dwarf satellite galaxies of the Milky Way: Fornax,
Leo~I and II, Sculptor, Sextans, Draco, Canes Venatici~I, and Ursa
Minor.  We use the published catalog of abundance measurements from
the previous paper in this series.  The measurements are based on
spectral synthesis of iron absorption lines.  For each MDF, we
determine maximum likelihood fits for Leaky Box, Pre-Enriched, and
Extra Gas (wherein the gas supply available for star formation
increases before it decreases to zero) analytic models of chemical
evolution.  Although the models are too simplistic to describe any MDF
in detail, a Leaky Box starting from zero metallicity gas fits none of
the galaxies except Canes Venatici~I well.  The MDFs of some galaxies,
particularly the more luminous ones, strongly prefer the Extra Gas
Model to the other models.  Only for Canes Venatici~I does the
Pre-Enriched Model fit significantly better than the Extra Gas Model.
The best-fit effective yields of the less luminous half of our galaxy
sample do not exceed $0.02~Z_{\sun}$, indicating that gas outflow is
important in the chemical evolution of the less luminous galaxies.  We
surmise that the ratio of the importance of gas infall to gas outflow
increases with galaxy luminosity.  Strong correlations of average
\feh\ and metallicity spread with luminosity support this hypothesis.
\end{abstract}


\section{Introduction}
\label{sec:intro}

The star formation history of a galaxy shapes the metallicity
distribution of its stars.  Therefore, simply counting the number of
stars in different bins of metallicity in a galaxy is a way to
quantify the gas dynamics during the history of star formation in the
galaxy.  How much gas was accreted by gravitational attraction?  How
much gas left the galaxy from supernova winds
\citep[e.g.,][]{dek86,gov10} or tidal or ram pressure stripping
\citep{lin83,mar03} from interaction with the Milky Way (MW)?

The most basic approach to answering these questions is to fit an
analytic model of chemical evolution to the observed metallicity
distribution function (MDF).  For example, one could assume that the
galaxy is a ``closed box'' \citep{tal71}.  In other words, the galaxy
begins its life with a fixed amount of gas.  It loses gas only to the
formation of stars, and it does not acquire new gas.  \citet*{van62}
and \citet{sch63} famously applied this model to the metallicity
distribution of G dwarfs in the solar neighborhood to find that the MW
disk is not a closed box.  Instead, it experiences more complicated
gas dynamics.  \citet{pag97} described some more complex analytic
models, including some that incorporate the accretion of external gas
during the lifetime of star formation.

Our aim is to examine the metallicity distributions of dwarf
spheroidal satellite galaxies (dSphs) of the MW in order to reveal how
gas infall and outflow affected their star formation histories.  We
use the \feh\ measurements for individual stars in eight MW dSphs from
\citeauthor*{kir10b} \citep{kir10b}.  The measurements are based on
spectral synthesis of iron lines from medium resolution spectra.
\citet*{kir08a} described the measurements in detail.  We modified the
procedure for determining \feh\ in Papers~I and II
\citep{kir09,kir10b}.

Several previous spectroscopic studies have examined MW dSph
metallicity distributions in the context of star formation history.
Most of them rely on a calibration between the summed equivalent
widths of the Ca infrared triplet and \feh\ \citep[e.g.,][]{rut97}.
\citet{tol01} first published a significant sample of Ca triplet
metallicities for the Sculptor and Fornax dSphs.  From the widths of
the metallicity distributions and the age spreads apparent from their
broadband colors combined with spectroscopic metallicities, they
concluded that both dSphs experienced extended star formation, unlike
globular clusters.  Since then, the Dwarf Abundances and Radial
Velocities Team (DART) have measured MDFs for Carina and Sextans in
addition to Fornax and Sculptor
\citep{tol04,hel06,bat06,bat08b,bat10}.  \citet{koc06} also conducted
their own Ca triplet survey of Carina.  \citet{kir08a} showed the
first spectral synthesis abundance measurements in dwarfs from
observations with a multi-object spectrograph.  They found extremely
metal-poor stars in the ultra-faint dSph sample of \citet{sim07}.
\citet{she09} also applied spectral synthesis to medium-resolution
spectra.  They obtained \feh\ measurements for 27 red giants in
Leo~II.  In \citeauthor*{kir09} \citep{kir09}, we presented the MDF
for the Sculptor dwarf galaxy based on the spectral synthesis of iron
lines for 388 red giants.

In this paper, we extend our analysis of Sculptor to seven additional
MW dSphs: Fornax, Leo~I and II, Sextans, Draco, Canes Venatici~I, and
Ursa Minor.  The sample of \ndsphstars\ stars in eight galaxies
permits a comparative look at star formation histories, particularly
as the properties of the MDFs change with dSph luminosity.
Furthermore, we examine how the distributions of \feh\ change as a
function of distance from the center of each dSph.

We describe three analytic chemical evolution models in
Sec.~\ref{sec:gce}.  They are Pristine, Pre-Enriched, and Extra Gas
Models.  In Sec.~\ref{sec:mdf}, we analyze the MDF of each dSph in
detail.  We discuss how each chemical evolution model may or may not
apply to different dSphs.  The results of previously published
numerical models \citep{lan03,lan04} are compared to the observed MDFs
in Sec.~\ref{sec:models}.  Section~\ref{sec:gradients} is devoted to
radial gradients and their relevance to the dwarf galaxies' star
formation histories.  We quantify trends of the MDF properties with
luminosity in Sec.~\ref{sec:trends} and the trends of the most likely
chemical evolution model parameters with dSph properties in
Sec.~\ref{sec:modeltrends}.  Finally, in Sec.~\ref{sec:conclusions},
we summarize our conclusions, point out shortcomings in the chemical
evolution models and our conclusions from them, and suggest how our
work may be improved in the future.


\section{Analytic Chemical Evolution Models}
\label{sec:gce}

The theory of galactic chemical evolution has progressed significantly
since \citet{tin80} codified the field.  \citet{lan04} applied
sophisticated multi-element models to the then scant spectroscopic
stellar abundance measurements in MW satellite galaxies.
\citet{mar06,mar08} created three-dimensional, hydrodynamic models of
isolated dwarf galaxies, but the sample size of observations was
inadequate to test their models.  Only recently, \citet{rev09} modeled
the abundance distributions of several MW satellites.  They obtained a
reasonable agreement between their predictions and the large samples
of the Dwarf Abundances and Radial Velocities Team's abundance
measurements in Fornax and Sculptor.

In \citeauthor*{kir09} of this series, we fit two different chemical
evolution models to the MDF of the Sculptor dwarf galaxy: a Simple
Model and an Extra Gas Model.  The Simple Model represented a leaky
box.  Although the gas was allowed to leave the galaxy, the galaxy
never acquired new gas.  Gas outflows do not affect the functional
form of the MDF as long as the nucleosynthetic yield $p$ is assumed to
be the effective yield and not the true metal yield of the stars.  In
the Simple Model, the initial gas of the galaxy was also allowed to be
be pre-enriched with a metallicity of $\mathfeh_0$.

We distinguish between the Simple Model with pristine initial gas
($\mathfeh_0 = -\infty$) and pre-enriched initial gas ($\mathfeh_0$ is
finite).  \citet{pag97} gave the functional form of the MDF of the
Pre-Enriched Model:

\begin{equation}
\frac{dN}{d\mathfeh} \propto \left(10^{\mathfeh} - 10^{\mathfeh_0}\right) \exp \left(-\frac{10^{\mathfeh}}{p}\right)\label{eq:preenriched}
\end{equation}

\noindent
where $p$ is in units of the solar metal fraction ($Z_\sun$).  For the
Pristine Model, one term vanishes:

\begin{equation}
\frac{dN}{d\mathfeh} \propto \left(10^{\mathfeh}\right) \exp \left(-\frac{10^{\mathfeh}}{p}\right) \: .\label{eq:pristine}
\end{equation}

In both the Pristine and Pre-Enriched Models, the peak of the MDF
increases with $p$.  However, a metal-rich dwarf galaxy did not
necessarily host supernovae with higher yields than a more metal-poor
galaxy, even if the Pristine Model is a good description for both
galaxies.  Because $p$ represents the effective yield, it encapsulates
both the supernova yield and gas outflow.  Larger supernova yield
increases $p$, and more intense gas outflow decreases $p$.  Therefore,
an equally valid interpretation of the hypothetical MDFs is that the
metal-rich (large $p$) galaxy retained gas more effectively than the
metal-poor galaxy.

The Extra Gas Model is the Best Accretion Model of \citet[][also see
  \citeauthor{pag97} \citeyear{pag97}]{lyn75}.  Unlike the Pristine
and Pre-Enriched Models, it allows the galaxy to access an additional
supply of gas available for forming stars during or between other
episodes of star formation.  For simplicity, we assume that the gas is
metal-free, and we assume that the rate of gas infusion decays over
time.  Furthermore, we assume a relation between the gas mass and the
stellar mass that permits an analytic solution to the differential
metallicity distribution.  \citeauthor{lyn75} generated such a
relation for which the gas mass reached a maximum and for which the
stellar mass rose asymptotically to its final value.  These
qualitative characteristics matched the simulations of \citet{lar74a}.
In this model, the gas mass $g$ in units of the initial gas mass is
related quadratically to the stellar mass $s$ in the same units:

\begin{equation}
g(s) = \left(1 - \frac{s}{M}\right)\left(1 + s - \frac{s}{M}\right)\label{eq:g}
\end{equation}

\noindent
where $M$ is a parameter greater than 1.  In the special case where
all of the gas is converted into stars and $p$ is the true yield, $M$
equals the final stellar mass in units of the initial mass of gas at
the onset of star formation.  When $M=1$, Eq.~\ref{eq:g} reduces to $g
= 1 - s$, which describes a closed box, wherein the gas mass is
depleted only by star formation.  Therefore, the Extra Gas Model
reduces to the Pristine Model when $M=1$.  Otherwise, an increase in
$M$ represents an increase in the amount of gas the galaxy gains.  The
following two equations describe the differential metallicity
distribution:

\begin{eqnarray}
\nonumber \mathfeh(s) &=& \log \bigg\{p \left(\frac{M}{1 + s - \frac{s}{M}}\right)^2 \times \\
            & & \left[\ln \frac{1}{1 - \frac{s}{M}} - \frac{s}{M} \left(1 - \frac{1}{M}\right)\right]\bigg\}\label{eq:s} \\
\nonumber \frac{dN}{d\mathfeh} &\propto&  \frac{10^{\mathfeh}}{p} \times \\
            & & \frac{1 + s\left(1 - \frac{1}{M}\right)}{\left(1 - \frac{s}{M}\right)^{-1} - 2 \left(1 - \frac{1}{M}\right) \times 10^{\mathfeh}/p}\label{eq:infall}
\end{eqnarray}

\noindent
Equation~\ref{eq:s} is transcendental, and it must be solved for $s$
numerically.  The solution to $s$ may then be put into
Equation~\ref{eq:infall}.

\citet{lyn75} named his model the Best Accretion Model because he
assumed that new gas became available for star formation by the
accretion or infall of gas onto the galaxy.  We have chosen the more
general name Extra Gas Model.  The model is too simplistic to
distinguish between different mechanisms of cold gas infusion.  For
example, the galaxy could contain hot gas.  If that gas cools between
episodes of star formation, then it becomes available to form stars.
However, the Extra Gas Model does require the newly available gas
supply to be metal-free, or at least much more metal-poor than the
bulk stellar metallicity at any given time.

Allowing for extra gas complicates the interpretation of the peak of
the MDF.  Both a larger supernova yield and smaller increase in the
supply of pristine gas would increase the peak of the MDF.  However,
gas outflow would still decrease the peak of the MDF.  Thankfully,
under the assumption of instantaneous mixing, the parameter $M$
uniquely quantifies the amount of extra gas, leaving $p$ to be
degenerate only between the supernova yield and gas outflow.

The Pristine, Pre-Enriched, and Extra Gas models all assume the
instantaneous recycling approximation (IRA) and the instantaneous
mixing approximation (IMA).  The IRA poorly reproduces the
distribution of secondary nuclides, which are produced on longer
timescales than primary nuclides.  Unfortunately, iron is a secondary
nuclide, but it is the most precisely measured of any element in
stellar spectroscopy because it has a large number of absorption lines
in the visible spectrum.  The IMA may not be appropriate for dSphs.
\citet{mar08} showed that inhomogeneous pollution from Type Ia SNe
affects the modeled MDF of a Draco-like dSph within two core radii.
However, analytic forms of the differential metallicity distributions
require the assumption of both the IRA and the IMA.


\section{Metallicity Distributions}
\label{sec:mdf}

We fit the Pristine, Pre-Enriched, and Extra Gas Models to the MDF of
each of the eight dSphs in the catalog from \citeauthor*{kir10b}.  The
model parameters---one for the Pristine Model and two for each of the
Pre-Enriched and Extra Gas Models---were determined by maximum
likelihood.  Each analytic metallicity distribution was treated as a
probability distribution ($dP/d\mathfeh$) which is normalized as
$\int_{-\infty}^{\infty} dP/d\mathfeh\: d\mathfeh= 1$.  The functional
forms of the probability distributions were identical to
Eqs.~\ref{eq:preenriched}, \ref{eq:pristine}, and \ref{eq:infall}.
The parameters were determined by maximizing the likelihood function
$L$.

\begin{eqnarray}
\nonumber L &=& \prod_i \int_{-\infty}^{\infty} \frac{dP}{d\mathfeh} \frac{1}{\sqrt{2 \pi}\,\delta\mathfeh_i} \times \\
            & & \exp \left(-\frac{\left(\mathfeh - \mathfeh_i \right)^2}{2\left(\delta\mathfeh_i\right)^2}\right) d\mathfeh \label{eq:lprod}
\end{eqnarray}

\noindent
The index $i$ represents each star in the observed MDF.  For
computational simplicity, the most likely parameters were actually
determined by minimizing the negative, logarithmic likelihood function
${\hat L}$.

\begin{eqnarray}
\nonumber {\hat L} &=& -\sum_i \ln \int_{-\infty}^{\infty} \frac{dP}{d\mathfeh} \frac{1}{\sqrt{2 \pi}\,\delta\mathfeh_i} \times \\
                   & & \exp \left(-\frac{\left(\mathfeh - \mathfeh_i \right)^2}{2\left(\delta\mathfeh_i\right)^2}\right) d\mathfeh \label{eq:lsum}
\end{eqnarray}

We initially estimated the model parameters that maximized likelihood
using the Powell optimization method.  Then we used a
Metropolis-Hastings Markov Chain Monte Carlo algorithm to refine the
fit and to estimate measurement uncertainties.  The proposal
distributions were normally distributed with $\sigma = 0.01$ for the
effective yield parameters and $\sigma = 0.1$ for the $\mathfeh_0$ and
$M$ parameters.  We conducted $10^3$ trials of the one-parameter
Pristine Model after $10^2$ burn-in trials.  We conducted $10^5$
trials of the two-parameter Pre-Enriched and Extra Gas Models after
$10^3$ burn-in trials.  The fiducial best-fitting parameters were
taken to be the median values of all of the trials.  Finally, we
computed the two-sided 68.3\% confidence interval.  The upper error
bar was the value that included 68.3\% of the trials above the median.
The lower error bar was the value that included 68.3\% of the trials
below the median.

The relative goodness of fit of one model over another may be
quantified by the ratios of the maximum likelihoods.  Once the
parameters for one model are determined by maximizing $L$, then the
model that better describes the data will have a larger maximum
likelihood, $L_{\rm max}$.  In the following sections, we quantify the
relative goodness of fit between two models as the logarithm of the
ratio of their maximum likelihoods: $\ln L_{\rm max}({\rm
  Model~A})/\allowbreak L_{\rm max}({\rm Model~B})$.

The Pristine Model consists of only one free parameter, the effective
yield.  Both the Pre-Enriched and Extra Gas Models depend on two free
parameters.  In fact, the Pristine Model is a special case of both
models ($\mathfeh_0 = -\infty$ for the Pre-Enriched Model and $M=1$
for the Extra Gas Model).  Therefore, the Pristine Model will never be
more likely than the Pre-Enriched or the Extra Gas Models.  (Sculptor
is an exception because we used the upper limit for
$\mathrm{[Fe/H]}_0$ in the calculation of $L_{\mathrm{max}}$ for the
Pre-Enriched Model.)  However, the ratio $\ln L_{\rm max}({\rm
  Pre\mbox{-}Enriched\ or\ Extra~Gas})/\allowbreak L_{\rm max}({\rm
  Pristine})$ may be close to zero, indicating that the extra free
parameter does not add significantly to the description of the MDF.

\begin{figure*}[p!]
\centering
\includegraphics[width=0.9\textwidth]{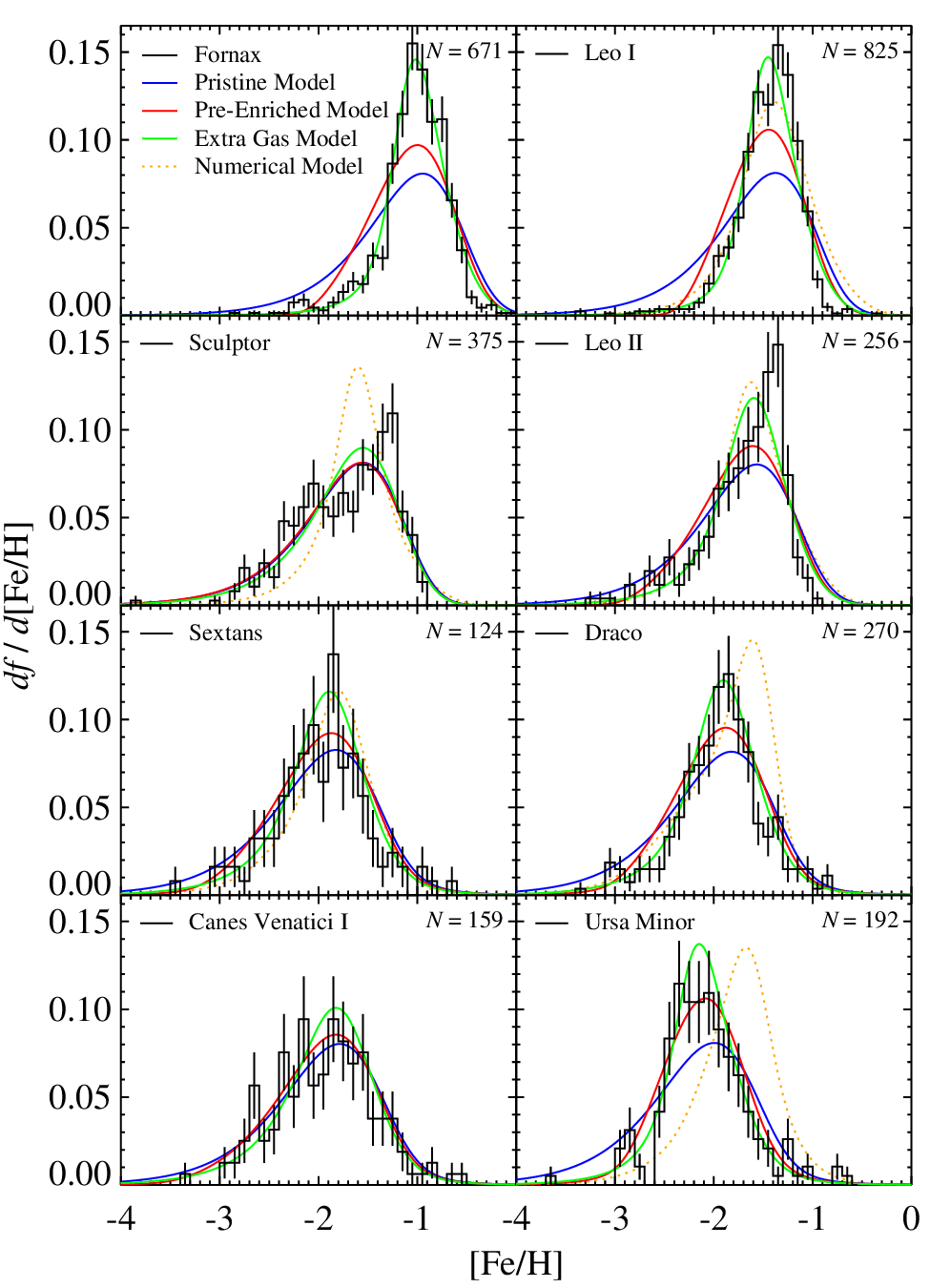}
\caption{The differential metallicity distribution in each dSph
  expressed as a fraction of the total number of observed stars.  The
  error bars represent Poisson counting statistics.  The panels are
  arranged from left to right and then top to bottom in decreasing
  order of dSph luminosity.  The black histograms show only stars with
  estimated uncertainties of $\delta\mathfeh < 0.5$.  The number of
  such stars is given in the upper right corner of each panel.  The
  blue, red, and green curves are the maximum likelihood fits to
  galactic chemical evolution models (Eqs.~\ref{eq:preenriched},
  \ref{eq:pristine}, and \ref{eq:infall}) convolved with the
  measurement uncertainties.  The dotted orange lines in some panels
  show predictions from numerical models (Sec.~\ref{sec:models})
  convolved with the measurement uncertainties.\label{fig:fehhists}}
\end{figure*}

Figure~\ref{fig:fehhists} shows the observed metallicity distributions
of each dSph with Poisson error bars.  The histograms include
measurements with estimated uncertainties on \feh\ less than 0.5~dex.
Stars with larger uncertainties are excluded from the figures for
clarity but not from the maximum likelihood fits.  This restriction
excludes fewer than 10 stars from the plots of the four most luminous
dSphs.  The lower stellar density in the four least luminous dSphs
necessitated targeting fainter stars.  Therefore, the fraction of
stars excluded from the plots is between 9\% and 12\% in Sextans,
Draco, Canes Venatici~I, and Ursa Minor.

In order to approximate the widening caused by measurement error, the
most likely analytic metallicity distributions have been smoothed.
The smoothing kernel was a sum of $N$ Gaussians, where $N$ is the
number of stars that passed the uncertainty cut.  The width
($\Delta\mathfeh$) of the $i^{\rm th}$ Gaussian corresponded to the
estimated uncertainty on the $i^{\rm th}$ measurement of \feh.  The
smoothing kernel was normalized to preserve the area under the MDF.
The kernel was constant with \feh\ because we have not observed a
significant correlation between our estimates of $\delta\mathfeh$ and
\feh\ except at the very lowest metallicities ($\mathfeh < -3$).

Our conclusions are valid for the subset of the stellar populations
observed.  Most dSphs have radial metallicity gradients (see
Sec.~\ref{sec:gradients}) such that the outermost stars are more
metal-poor than the innermost stars.  Our spectroscopic observations
were centered on the dSphs in order to maximize the number of member
stars.  Consequently, our samples are more metal-rich and probably
younger than the dSphs' entire stellar populations.  This effect is
especially pertinent because some dSphs have been known to lose their
outermost, oldest stars via tidal stripping by the MW \citep[e.g.,
  Carina,][]{maj00b}.  Consequently, our results are not applicable to
the entire star formation histories of the dSphs, such as Leo~I
\citep{soh07}, that may have shed significant fractions of their
older, more metal-poor stars.

\subsection{Fornax}
\label{sec:for}

Fornax is the most luminous of the dSphs that we consider, and it the
most luminous intact dSph that orbits the MW.  Sagittarius is more
luminous, but the MW has tidally disrupted it \citep{iba94}, and it
may have been too luminous \citep{nie10} and too disky \citep{pen10}
to belong to the same class of galaxies as the surviving dwarf
spheroidals.  In agreement with the metallicity-luminosity relation
for MW dwarf satellites \citep{mat98,kir08b}, Fornax also shows the
highest peak \feh\ of all of the MDFs in Fig.~\ref{fig:fehhists}.  It
has a median \feh\ of $\forfehmedian$.

The Pristine and Pre-Enriched Models of chemical evolution poorly
match the observed MDF.  They do not allow a narrow peak, nor do they
allow a sharp change in slope on the metal-poor end of the peak.
Instead, the Extra Gas Model matches the observed MDF much better.
The logarithm of the ratios of the maximum likelihood for the Extra
Gas Model to the maximum likelihoods for the Pristine and Pre-Enriched
models are $\ln L_{\rm max}({\rm Extra~Gas})/\allowbreak L_{\rm
  max}({\rm Pristine}) = \forprobratio$ and $\ln L_{\rm max}({\rm
  Extra~Gas})/\allowbreak L_{\rm max}({\rm Pre\mbox{-}Enriched}) =
\forprobratioinfallpre$.  The most likely $M$ parameter is
$\forminfall_{-{\forminfallerrlo}}^{+{\forminfallerrhi}}$, indicating
a large departure from the Pristine Model.  The fraction of stars
formed from gas that fell into the system is $(M-1)/M$, or $\sim 87\%$
in this case.  The Extra Gas Model does very well at matching the
symmetric peak, but the narrowness of the peak demands a large $M$,
which in turn causes an underestimate of the frequency of
low-metallicity stars and an overestimate of the frequency of
high-metallicity stars.  If we have overestimated the measurement
uncertainties on \feh, then the model's intrinsic, unconvolved peak
could be wider, and $M$ could be smaller.  That solution would fit the
metal-poor tail and the steep metal-rich slope better.

The reasonable fit of the Extra Gas Model suggests the following
extended star formation history for Fornax.  Fornax began as a dark
matter subhalo with a gas mass of $\sim 13\%$ of its final stellar
mass.  Pristine, zero-metallicity gas fell in gradually.  The first
stars in Fornax formed from the initial gas.  The most massive of
these stars exploded very quickly and enriched the interstellar medium
(ISM) rapidly.  The less massive, long-lived stars from early in
Fornax's history still populate the metal-poor tail today.  As more
gas fell in, star formation continued from the enriched gas.  Because
the timescale for Type~II SN metal enrichment
is short, most of the stars in Fornax formed after several generations
of massive stars already enriched the ISM.  As a result, most Fornax
stars occupy the relatively metal-rich peak.  This picture is
qualitatively consistent with the conclusions of \citet{orb08}.  From
color-magnitude diagrams, they deduced that 27\% of Fornax stars are
older than 10~Gyr, and 33\% are younger than 5~Gyr.  The old stars
enriched the ISM for the majority population---intermediate-age and
younger.

\citet{bat06} also observed a spectroscopic (Ca triplet-based) MDF of
Fornax.  Their sample extended beyond the tidal radius at 1.2~deg
\citep{irw95}.  Our sample, which has a higher spatial density,
extends to 16.3~arcmin.  \citeauthor{bat06}'s MDF within 24~arcmin
shows the same qualitative shape as our MDF.  The peak of their MDF is
$\sim 0.15$~dex lower than the peak we measure, but they showed that
Fornax's \feh\ decreases beyond the radial extent of our data.
Therefore, we expect that their more extended sample would be more
metal-poor.  In fact, they argued for the importance of gas outflows
and infall for the star formation history of Fornax based not on the
MDF but on the radial metallicity gradient and non-equilibrium
kinematics.

\subsection{Leo~I}
\label{sec:leoi}

Leo~I is the second most luminous intact dSph that orbits the MW.
Accordingly, its peak \feh\ is lower than that of Fornax, with a
median \feh\ of $\leoifehmedian$ and an intrinsic dispersion of
0.36~dex.  In a Ca triplet study, \citet{gil09} found that the mean
metallicity is $\rm{[M/H]} = -1.2$ with a dispersion of 0.2~dex.  The
shape of the MDF resembles that of Fornax.  The MDF has a narrow peak
and a long metal-poor tail.  The most likely extra gas parameter is $M
= \leoiminfall_{-{\leoiminfallerrlo}}^{+{\leoiminfallerrhi}}$.  The
presumed increase in gas supply was about as intense for Leo~I as for
Fornax.  Again, the Pristine and Pre-Enriched Models are not good
descriptions of the MDF: $\ln L_{\rm max}({\rm Extra~Gas})/\allowbreak
L_{\rm max}({\rm Pristine}) = \leoiprobratio$ and $\ln L_{\rm
  max}({\rm Extra~Gas})/\allowbreak L_{\rm max}({\rm
  Pre\mbox{-}Enriched}) = \leoiprobratioinfallpre$.  In terms of the
star formation history proposed for Fornax, it seems that Leo~I
encountered about the same history, with a gradual increase in the gas
supply that formed a small fraction of very metal-poor stars.

Leo~I has the largest distance \citep[254~kpc,][]{bel04} among the
``classical'' dSphs.  If ram pressure stripping of gas plays a role in
the star formation history of dwarf galaxies
\citep[e.g.,][]{lin83,mar03}, then Leo~I may have experienced less gas
stripping than Sculptor or Fornax.  Its highly elliptical orbit
\citep{soh07,mat08} allows Leo~I to spend most of its time far from
the MW center, allowing little time for strong tidal interactions.
(When it does pass through the disk, however, it passes at high
velocity.)  In its lifetime, Leo~I experienced a close perigalacticon
just once \citep{mat08} or a few times \citep{soh07}.  The possibly
less intense interaction with the MW could have preserved enough gas
in Leo~I for it to appear more like Fornax than Sculptor.  The star
formation histories deduced by \citet{orb08} support that idea.  Their
measured mean ages are 6--8~Gyr for Leo~I and Fornax stars.  However,
\citet{sme09} interpreted their Hubble Space Telescope Advanced Camera
for Surveys (HST/ACS) imaging as evidence that the dominant population
of Leo~I is older than 10~Gyr.  Our MDF gives only a qualitative
suggestion that Leo~I has experienced an extended star formation
history, like Fornax, but the MDF alone is insufficient to quantify
the mean stellar age.

The close perigalactic passes probably caused Leo~I to lose stars in
addition to gas.  The stars most susceptible to tidal stripping are
the outermost ones, which tend to be older and more metal-poor (see
Sec.~\ref{sec:gradients}).  Therefore, the observed MDF represents
only the stars that Leo~I still retains but not the complete star
formation history of the galaxy.

\citet{bos07} also measured the MDF for Leo~I.  They used their own
calibration between the Ca triplet equivalent width and metallicity.
They preferred to calibrate to [Ca/H], but they also derived MDFs in
terms of \feh.  The average [Ca/H] of their MDF $-1.34$.  The average
\feh\ of our MDF is $\leoifehmean$.  \citeauthor{bos07}'s sample
reached a radial extent of at least 20 arcmin.  The maximum extent of
our sample is 14.4~arcmin, and it becomes sparsely sampled beyond
8~arcmin.  The shape of \citeauthor{bos07}'s MDF is slightly
asymmetric, with a metal-poor tail.  Our MDF is highly asymmetric.
The sparser sampling of their MDF (102 stars) compared to ours
(\leoinunique\ stars) and the different measurement techniques may
explain the different appearances.

\subsection{Sculptor}
\label{sec:scl}

We have already shown and analyzed the MDF for Sculptor
(\citeauthor*{kir09}).  We now place the MDF for Sculptor in the
context of the MDFs for the other seven dSphs.  Sculptor is the third
most luminous intact dSph that orbits the MW.  Its MDF might be
expected to appear similar to Fornax and Leo~I.  However, Sculptor's
MDF is unlike any of the other seven dwarfs.  None of the three models
adequately describe the MDF.

First, the Pristine Model is too narrow to reproduce the broad
\feh\ distribution, including a possible secondary peak at $\mathfeh =
-2.1$.  Second, the Pre-Enriched Model might have been able to attain
a better-fitting shape with a high enough $\mathfeh_0$.  However, star
S1020549 at $\mathfeh = -3.87 \pm 0.21$, confirmed with
high-resolution spectroscopy by \citet*{fre10a}, demands a low initial
enrichment.  The upper limit on the initial enrichment that we derive
is $\mathfeh_0 < \sclfehpre$.  Third, the most likely Extra Gas Model,
with $M = \sclminfall_{-{\sclminfallerrlo}}^{+{\sclminfallerrhi}}$, is
nearly identical to the Pristine model.  Increasing $M$ would only
narrow the MDF further.

A more proper nucleosynthetic treatment of iron would also better
reproduce the MDF.  Our analytic models assume instantaneous
recycling, meaning that they do not incorporate a delay time between
the births of stars and the return of enriched material into the ISM.
In \citeauthor*{kir10a} \citep{kir10a}, we show that a numerical model
that relaxes the instantaneous recycling approximation results in a
model MDF that better fits Sculptor's observed MDF.  However, even the
more sophisticated model does not reproduce the apparent bimodality in
the observed MDF (two peaks at $\mathfeh = -2.1$ and $-1.3$).

The kinematic distribution of Sculptor's stars may provide some
insight on the bimodal MDF.  \citet{tol04} found that a two-component
model best describes Sculptor.  The stars separate into a centrally
concentrated, metal-rich ($\mathfeh > -1.7$), kinematically cold
($\sigma_v = 7$~km/s) component and an extended, metal-poor ($\mathfeh
< -1.7$), kinematically warm ($\sigma_v = 11$~km/s) component.  A
two-component model would do a much better job of representing our
observed MDF of Sculptor.  Combining any two of the three chemical
evolution models with different peaks in \feh\ would generate a broad
\feh\ distribution.  However, we reserve a two-component analysis for
a study that includes kinematic data.

The shape of the Sculptor MDF is very different from Fornax.
\citet{dek86} suggested that galactic outflows play a large role in
dSph formation.  If Sculptor is less massive than Fornax, then winds
from supernovae could have rapidly depleted Sculptor of the enriched
gas necessary to create the more metal-rich stars.  Therefore, the
smooth shape of Sculptor's MDF compared to the peaked shape of
Fornax's MDF may indicate that galactic outflows were more important
than an increase in the gas supply in Sculptor's history, whereas the
reverse was true for Fornax.

\subsection{Leo II}
\label{sec:leoii}

Leo~II, Sextans, and Carina are the next most luminous dSphs that
orbit the MW.  Their luminosities are nearly indistinguishable within
the error bars given by \citet{irw95}.  Leo~II continues the trend
established by Fornax and Leo~I.  It has a slightly lower median
\feh\ ($\leoiifehmedian$) than that of Leo~I.  In accordance with its
lower luminosity, its extra gas parameter $M =
\leoiiminfall_{-{\leoiiminfallerrlo}}^{+{\leoiiminfallerrhi}}$ is also
lower than that of Leo~I.  The Pristine and Pre-Enriched models are
still not good fits, but they are better representations of the MDF
than for Fornax and Leo~I, where the Extra Gas Model departs more
severely from the Pristine Model ($M > 7$).

None of the models can reproduce the steep slope on the metal-rich
side of the peak of Leo~II.  One possible solution to the abrupt drop
in the frequency of metal-rich stars is a terminal wind.  For example,
a Closed Box Model that is truncated at some metallicity ($\mathfeh
\sim -1.3$ in this case) might produce an MDF similar to that of
Leo~II.  \citet{win03}, among others, considered such a model in her
description of the MDFs of Sculptor, Draco, and Ursa Minor.

\citet{orb08} reported similar mass-weighted average stellar ages for
Fornax (7.4~Gyr), Leo~I (6.4~Gyr), and Leo~II (8.8~Gyr).  They listed
the mean stellar ages for Sculptor, Sextans, Ursa Minor, and Draco as
larger than 10~Gyr, and \citet{mart08b} gave a mean age larger than
10~Gyr for Canes Venatici~I.  It is not surprising, then, that Fornax,
Leo~I, and Leo~II---the younger dSphs---have similar MDFs.
Furthermore, their younger mean ages may have allowed them extra time
to accrete gas while they were forming stars.  The increased supply of
gas would have caused their peaks to be narrower than can be explained
by a Leaky Box Model and would explain why the Extra Gas Model fits
the best of the three chemical evolution models that we consider.

Finally, the distance to Leo~II is nearly as great as Leo~I.
\citet{sie10} gave a distance of 219~kpc.  Although the orbit of
Leo~II is unknown, it could spend much of its time far outside of the
range of disruptive gravitational interaction with the MW.  It could
have been spared gas stripping, which other dSphs like Sculptor or
Sextans may have encountered.  However, if Leo~II does spend most of
its time in a low density region of the Local Group, then it would
likely not encounter the gas reservoir required to explain our
interpretation of the MDF as indicative of an increase in the gas
reservoir.

\citet{bos07} observed the MDF of Leo~II in addition to Leo~I.  They
measured a mean [Ca/H] of $-1.65$, whereas we measure a mean \feh\ of
$\leoiifehmean$.  The radial extent of their survey was nearly the
same as our survey.  The shape of the Ca triplet MDF is at least
qualitatively similar to our MDF.  Both show an abrupt fall-off in the
frequency of metal-rich stars.  \citeauthor{bos07}\ interpreted the
absence of metal-rich stars as evidence for rapid gas loss.  Indeed,
the effective metallicity yield that we measure ($\leoiiyieldinfall
\pm \leoiiyieldinfallerr~Z_{\sun}$) is too low to be explained by
completely retained SN ejecta.  The galaxy must have lost some gas,
but the shape of the MDF also mandates that it accreted
low-metallicity gas.

\subsection{Sextans}
\label{sec:sex}

Despite its similar luminosity to Leo~II, Sextans displays a
differently shaped MDF.  It is more symmetric, with a shallower slope
on the metal-rich side of the peak.  Most strikingly, the median
\feh\ for Sextans ($\sexfehmedian$) is significantly lower than the
median \feh\ for Leo~II ($\leoiifehmedian$).  As a result, the most
likely yield for the Extra Gas Model is much lower in Sextans
($\sexyieldinfall \pm \sexyieldinfallerr~Z_{\sun}$) than in Leo~II
($\leoiiyieldinfall \pm \leoiiyieldinfallerr~Z_{\sun}$) despite
Sextans's similar extra gas parameter, $M =
\sexminfall_{-{\sexminfallerrlo}}^{+{\sexminfallerrhi}}$.  However,
the Extra Gas Model may not be the best description of the MDF.
Quantitatively, neither the Pre-Enriched nor the Extra Gas Model is
significantly preferred: $\ln L_{\rm max}({\rm
  Pre\mbox{-}Enriched})/L_{\rm max}({\rm Extra~Gas}) =
\sexprobratiopreinfall$.

Sextans has a very large tidal radius \citep[$160 \pm
  50$~arcmin,][]{irw95}, which means that a complete MDF requires
extensive radial sampling.  \citet{bat10} observed the Ca
triplet-based MDF of Sextans to very large distance from the dSph
center.  Their sample extended to $1.8\arcdeg$ whereas our sample
extends to only 21.4~arcmin.  Their more complete MDF is more
metal-poor than ours because they detected a radial gradient of
$-0.33~\mathrm{dex~kpc}^{-1}$.  Our sample is too centrally
concentrated to detect a gradient.  \citeauthor{bat10}\ also found
that the shape of the MDF changes as a function of radius.  Sextans
appears to have two metallicity populations: a metal-rich population
within $0.8\arcdeg$ and a metal-poor population beyond $0.8\arcdeg$.
Therefore, our results should be interpreted as applicable only to the
innermost population, which presumably formed more recently than the
outer population.

\subsection{Draco}
\label{sec:dra}

Draco, Canes Venatici~I, and Ursa Minor form the next group of dSphs
with indistinguishable luminosities.  The MDFs of these three dSphs
and Sextans are broadly related.  Their mean \feh\ values all lie
between $-2.2$ and $-1.9$, and their MDFs seem to be more symmetric
than the more luminous dSphs.

The MDF of Draco is similar to that of Sextans.  Most of the shape
parameters (mean, median, dispersion) for the two MDFs are nearly
identical.  In particular, the observed MDF is more peaked than the
Pristine Model, and it has a metal-poor tail.  Formally, the Extra Gas
Model is a better fit to Draco than the other models ($\ln L_{\rm
  max}({\rm Extra~Gas})/L_{\rm max}({\rm Pristine}) = \draprobratio$
and $\ln L_{\rm max}({\rm Extra~Gas})/L_{\rm max}({\rm
  Pre\mbox{-}Enriched}) = \draprobratioinfallpre$).  On the other
hand, the Pre-Enriched Model fits Sextans slightly better than the
Extra Gas Model.  However, Draco is better sampled than Sextans, with
more than twice as many stars with measurements of \feh.  As a result,
the most likely chemical evolution parameters for Draco are more
secure than for Sextans.

\citet{win03} also observed and modeled the MDFs for Draco, Sculptor,
and Ursa Minor.  Her analytic models included abrupt or continuous gas
loss with initially pristine or pre-enriched gas.  These
single-component models did not fit any of the three MDFs very well,
but a two-component model for Draco did work well.  The two components
were two Leaky Box Models with pre-enrichment.  The models reproduced
the Draco MDF shape very well.  However, the shape of
\citeauthor{win03}'s Draco MDF was different from ours.  Her MDF had
two peaks.  The different shape results from different measurement
techniques and different radial sampling (a maximum of $\sim 30$
arcmin for \citeauthor{win03}'s and 60~arcmin for our sample).
Nonetheless, the two-component experiment demonstrates that an
accurate description of some MDFs (especially Sculptor) may require
multi-component models.

\subsection{Canes Venatici I}
\label{sec:cvni}

The MDF of Canes Venatici~I resembles a normal distribution more than
any of the other seven dwarfs.  Although it has a slight metal-poor
tail, it is nearly symmetric.  Canes Venatici~I shows the least
preference for the Extra Gas Model of all the dSphs in
Fig.~\ref{fig:fehhists} except for Sculptor.  The most likely extra
gas parameter is $M =
\cvniminfall_{-{\cvniminfallerrlo}}^{+{\cvniminfallerrhi}}$, and the
preference of the Extra Gas Model over the Pristine Model is slight
($\ln L_{\rm max}({\rm Extra~Gas})/L_{\rm max}({\rm Pristine}) =
\cvniprobratio$).

Overall, the MDF of Canes Venatici~I fits a Leaky Box Model best of
all the dSphs shown here.  It is the only dSph in our sample that fits
the Pre-Enriched Model significantly better than the Extra Gas Model
($\ln L_{\rm max}({\rm Pre\mbox{-}Enriched})/L_{\rm max}({\rm
  Extra~Gas}) = \cvniprobratiopreinfall$).  The dSph's large distance
\citep[210~kpc,][]{kue08} would make it less susceptible to gas
accretion from the MW than the other dSphs in its luminosity class,
Draco and Ursa Minor.  Continuous gas outflow---from SN winds, for
example---would not drastically affect the shape of the MDF.  It would
instead decrease the effective yield $p$ and cause a low peak \feh,
which is observed.

\subsection{Ursa Minor}
\label{sec:umi}

The luminosity of Ursa Minor is within a factor of two of Sextans.
\citet{orb08} derived identical star formation histories for both
dSphs: no stars younger than 12~Gyr.  Ursa Minor, with the slightly
lower luminosity, has a correspondingly lower median \feh:
$\umifehmedian$ compared to $\sexfehmedian$.  The most likely Extra
Gas Model indicates intense gas inflow over the lifetime of star
formation, despite the lack of an obvious G dwarf problem.  The $M$
parameter for the Extra Gas Model is
$\umiminfall_{-{\umiminfallerrlo}}^{+{\umiminfallerrhi}}$ for Ursa
Minor compared to
$\sexminfall_{-{\sexminfallerrlo}}^{+{\sexminfallerrhi}}$ for Sextans.

None of the models accurately reproduce the sudden absence of
metal-poor stars at $\mathfeh < -3$.  \citet{hel06} invoked the
Pre-Enriched Model as a possible explanation for the apparent dearth
of metal-poor stars in Fornax, Sculptor, Carina, and Sextans.  Since
then, extremely metal-poor stars have been discovered in dSphs
\citep{kir08b,kir09,geh09,coh09,coh10,fre10a,fre10b,sim10a,sim10b,nor10a,nor10b,sta10,taf10}.
These discoveries do not preclude pre-enrichment in all dSphs.  In
fact, the most likely initial metallicity for the Pre-Enriched Model
for Ursa Minor is $\mathfeh_0 =
\umifehpre_{-{\umifehpreerrlo}}^{+{\umifehpreerrhi}}$.  However,
\citet{coh10} have discovered two stars with $\mathfeh < -3$ in Ursa
Minor using high-resolution spectra.  Our sample includes seven such
stars, including one with $\mathfeh = -3.62 \pm 0.35$.

On closer inspection, Ursa Minor is an outlier from its luminosity
class.  Whereas all of the other dSphs show at least a hint of a
metal-poor tail, Ursa Minor shows a metal-rich tail.  Ursa Minor's
large negative radial velocity ($\umimeanv$~km~s$^{-1}$) rules out
significant contamination from metal-rich Galactic stars.  The
Pristine Model can explain neither the absence of a metal-poor tail
nor the existence of a metal-rich tail.  The most likely Pre-Enriched
Model is too symmetric.  The model with the sharpest peak is the Extra
Gas Model with a very large $M$ and a very low yield.  The most likely
parameters are $M =
\umiminfall_{-{\umiminfallerrlo}}^{+{\umiminfallerrhi}}$ and $p =
\umiyieldinfall \pm \umiyieldinfallerr~Z_{\sun}$, the most extreme of
all eight dSphs.


\section{Numerical Chemical Evolution Models}
\label{sec:models}

We compare the observed MDFs of six galaxies (Leo~I, Sculptor, Leo~II,
Sextans, Draco, and Ursa Minor) to predictions of detailed numerical
models of chemical evolution in addition to the analytic models.
Figure~\ref{fig:fehhists} includes the MDFs from the numerical models
as dotted lines.  The model MDFs have been convolved with the same
function as the analytic models to approximate observational
uncertainty (Sec.~\ref{sec:mdf}).  In the adopted numerical models
\citep{lan03,lan04}, the evolution of each galaxy is mainly controlled
by the assumptions regarding the star formation (SF) history, SF
efficiency, and galactic wind.  All models adopt an infall of pristine
gas, low star formation rate (SFR), and high galactic wind efficiency.
The low SFR and the high efficiency of the wind give rise to a peak in
the MDFs at low \feh\ (approximately between $-1.6$ to $-2.0$) whereas
the long infall timescale of pristine gas allows the models to form a
low number of metal-poor stars, similar to the observed frequency.

In contrast to the analytic models, the numerical models were not
adjusted to fit the present data.  The predictions are the same as in
previous papers, in which the models were adjusted to better match
different observational data.  The numerical models for Ursa Minor and
Draco \citep{lan07} were adjusted to approximate the observed MDFs
based on photometric metallicities \citep{bel02}, whereas the models
for Leo~I and II \citep{lan10} were calibrated to match the
\feh\ distribution inferred from Ca triplet lines
\citep{koc07a,koc07b,bos07,gil09}.  The predictions for Sextans and
Sculptor are true predictions; they have not been calibrated to match
any observational data because they were published before data were
available \citep{lan04}.  We can use the comparisons to infer what
modifications should be made in the models, especially those regarding
SF and wind efficiencies.

The differences in shape between the predicted and observed MDFs may
be described in terms of how the modeled SF history could be changed
to achieve a better fit.  Higher SFRs seem to be necessary in the case
of Leo~I and Leo~II because these two models exhibit MDF peaks $\sim
0.4$~dex lower than observed.  The model of Leo~I is characterized by
two long episodes of SF at 14~Gyr and 9~Gyr ago, lasting 5~Gyr and
7~Gyr respectively, with a low efficiency ($\nu \simeq
0.6~\rm{Gyr}^{-1}$) and by the occurrence of a very intense galactic
wind with a rate 9 times higher than the SFR ($w_i = 9$).  On the
other hand, the Leo~II model adopts lower SF and wind efficiencies
($\nu \simeq 0.3~\rm{Gyr}^{-1}$ and $w_i = 8$) and just one long
episode of SF at 14~Gyr ago, lasting 7~Gyr.  Modifying the duration
and epoch of the SF episodes would not change the predictions
substantially, whereas the wind efficiency influences the position of
the peak in the MDFs and, most significantly, the relative number of
metal-rich stars.  Because the shapes of the predicted MDFs are
similar to the observations, SF efficiency is the only parameter that
requires adjustment.  Increasing $\nu$ might lead to a better fit in
both cases.

For Ursa Minor and Draco, the predicted MDFs are 0.3--0.5~dex more
metal-rich than the observed MDFs.  These two galaxies are
characterized by short, older periods of SF (4~Gyr and 3~Gyr,
respectively) compared to the other dSphs and by the lowest SF
efficiencies ($\nu \simeq 0.1~\rm{Gyr}^{-1}$ and $\nu \simeq
0.05~\rm{Gyr}^{-1}$, respectively) among the six models analyzed.  It
seems that these values need to be further decreased to match the
observations, especially for Ursa Minor.  This galaxy also exhibits a
more extended high metallicity tail, which could be reproduced by a
lower galactic wind efficiency.  To prevent the subsequent increase in
the peak \feh, the SFR should also be decreased.  In contrast to Ursa
Minor, the prediction for Draco fits the number of metal-poor stars
well but overestimates the peak of the MDF and the number of high
metallicity stars.  The discrepancies are probably consequences of low
wind efficiency.  A higher wind efficiency would decrease the SFR
after the onset of the wind, lowering the number of metal-rich stars
born.  Such a model would better match the data by creating an MDF
with a peak at lower \feh\ and with fewer metal-rich stars.

The predicted MDF of Sextans reproduces the observed data well.  The
model adopts a long episode of SF (longer than 4~Gyr) with low rates
and an intense galactic wind ($\nu \simeq 0.08~\rm{Gyr}^{-1}$ and $w_i
= 9$), giving rise to a main stellar population with low \feh\ ($\sim
-1.8$~dex).  There seems to be an underprediction of the frequency of
metal-poor stars, probably due to the extended infall of gas and to
the prolonged SF.

For Sculptor, as in the case of analytic models, the prediction does
not fit the observed MDF.  The observations show a distribution much
broader than the predictions.  As discussed in Sec.~\ref{sec:scl}, the
cause of the broad MDF could be two different stellar population with
different metallicities and spatial extents.  \citeauthor{lan10}'s
model cannot separate the two different populations because it uses
only one zone.  Instead, the different populations may be the result
of different SFRs in the central and in the outer regions, perhaps due
to different gas densities.  The result would be different chemical
enrichment and different mean metallicities in each region.  We
explore this idea more generally in the next section.


\section{Radial Metallicity Distributions}
\label{sec:gradients}

Galaxies often show radial metallicity gradients
\citep[e.g.,][]{meh03}.  We discuss three processes that may be
responsible for gradients.  (1) The star formation rate in a galaxy
depends on the gas density \citep{sch59,ken83}, and the gas density
increases toward the bottom of the gravitational potential well in the
center of the galaxy.  Therefore, the center of the galaxy may
experience the highest star formation intensity and consequently may
show the highest mean metallicity.  (2) MW satellite galaxies may lose
gas through tidal or ram pressure stripping \citep{lin83} from the MW.
The gas that lies far from the center is more loosely bound to the
dSph than gas that lies at the center.  Therefore, gas leaves the dSph
from the outside in.  For dSphs affected by gas loss, later episodes
of star formation occur closer to the center of the galaxy.  Because
later episodes of star formation occur from more metal-rich gas, gas
loss creates a stellar metallicity gradient, with the center of the
dSph being more metal-rich.  This scenario requires star formation to
occur during the process of gas loss.  (3) The angular momentum of
accreted, low-metallicity gas will not allow the gas to reach the
center of the galaxy.  If stars form from this gas, they will be
metal-poor and mostly confined to large radius.  Therefore, gas
accretion can also generate radial metallicity gradients.

\begin{figure}[t!]
\centering
\includegraphics[width=\columnwidth]{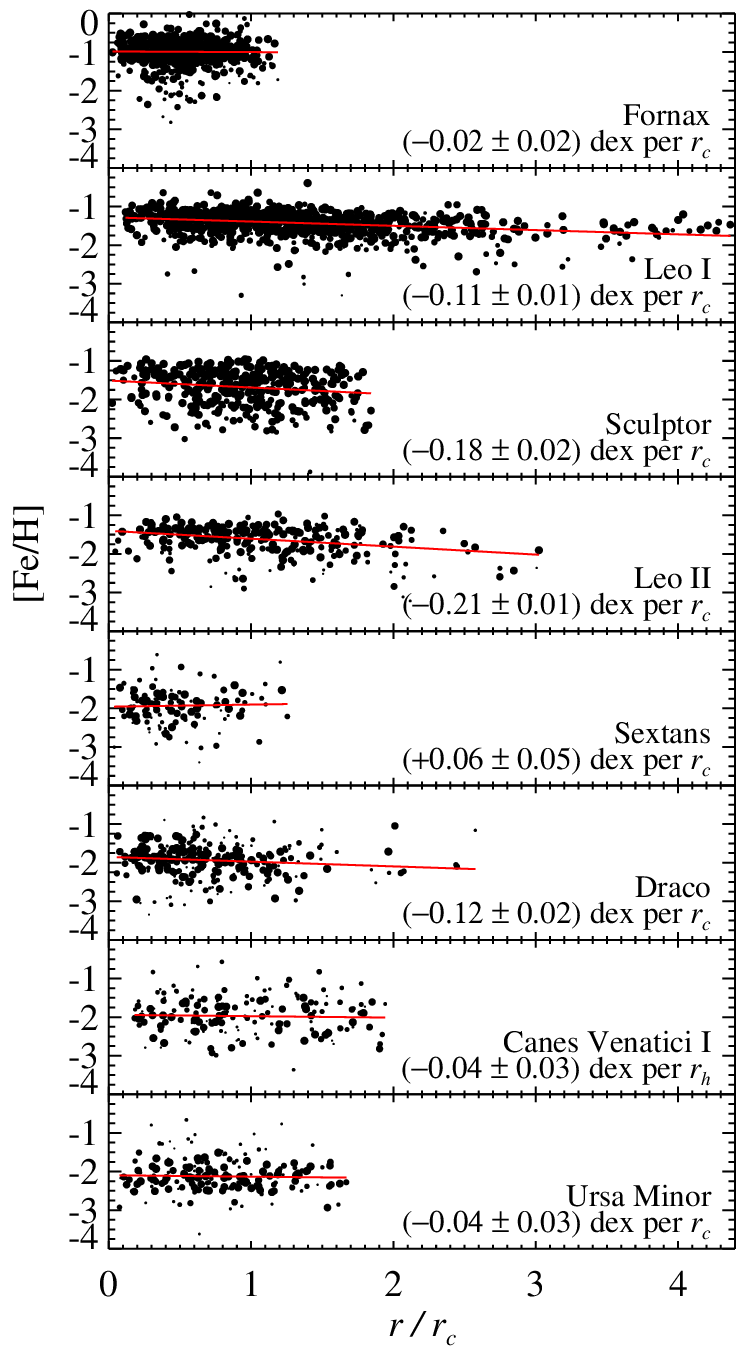}
\caption{Radial distributions of \feh\ as a function of radius in
  units of the core radius \citep[$r_c$,][]{irw95} or half-light
  radius \citep[$r_h$,][]{mart08b} in the case of Canes Venatici~I.
  The dSphs are arranged in decreasing order of luminosity.  The red
  line is the least squares linear fit, whose slope is given in each
  panel.  The point sizes are inversely proportional to the
  uncertainty in \feh.\label{fig:gradients}}
\end{figure}

\input{table_gradients}

Figure~\ref{fig:gradients} shows the radial distributions of \feh.
The $x$-axis gives the distance from the dSph center in units of the
core radius from a King profile fit \citep{irw95} or half-light radius
for Canes Venatici~I \citep{mart08b}.  The lines are the least squares
linear fits.  Table~\ref{tab:gradients} gives the slopes of these
lines in terms of angular distance from the dSph center in degrees,
projected distance from the dSph in kpc, and projected distance in
core radii.  Four dSphs, Leo~I, Sculptor, Leo~II, and Draco, show
significantly negative slopes.  The slopes of the other four dSphs are
consistent with zero or nearly consistent with zero.

The eight dSphs in our sample are insulated against Galactic
contamination by their Galactic latitudes ($|b| > 40\arcdeg$), radial
velocities ($|v_{\rm helio}| > 200~\rm{km~s}^{-1}$), or both.  Based
on the velocity distribution of stars excluded by radial velocity, we
estimate that fewer than 5\% of the stars are Galactic contaminants
even in the worst cases (Sextans and Draco).  However, some Galactic
halo stars may still contaminate our samples.  \citet{sch09} show that
the local Galactic halo MDF is broad with a peak at $\mathfeh = -2.1$
and a secondary peak at $\mathfeh = -1.1$.  We would not necessarily
measure the correct metallicity for these stars because we assume the
same distance modulus for all stars in a given dSph sample.  Halo
stars could lie in front of or behind the dSphs.  The distance modulus
affects the metallicity measurements through the photometric
determination of surface gravity and partially photometric
determination of effective temperature.  We estimate that the
intrinsic breadth of the halo MDF coupled with the increased
measurement errors would cause halo stars to appear uniformly
distributed in the range $-3 \la \mathfeh \la -1$.  Fornax and Leo~I
contain many stars at higher metallicities.  Halo contamination would
cause us to infer a more negative metallicity gradient for those two
dSphs because the ratio of dSph to halo stars decreases with radius.
However, the radial velocity distributions of stars in the Fornax and
Leo~I slitmasks suggest contamination at the level of 1\% or less.
Contamination in the other dSphs would tend to flatten the measured
gradients and reduce their significance because the contaminating
population has roughly the same [Fe/H] distribution as the average
distribution for the entire dSph.  However, we reiterate that the
contamination is small.

We might have expected that Fornax has a strong radial gradient due to
its longer star formation history \citep{orb08}, with successive
generations becoming more and more centrally concentrated as the gas
was depleted from the outskirts.  However, the slope of the best-fit
line is consistent with zero.  \citet{bat06} did find a radial
metallicity gradient in Fornax.  Their sample extended beyond the
tidal radius whereas our sample reaches to only about the core radius.
Within the radial bounds of our sample, \citeauthor{bat06}'s
measurements do not show evidence for a radial gradient, either.
Therefore, our sample is too limited in angular extent to draw
definitive conclusions on the presence of a radial metallicity
gradient.

Sextans, Canes Venatici~I, and Ursa Minor also show insignificant
slopes of \feh\ with radius within our centrally restricted samples.
The lack of gradients possibly indicates short star formation
durations.  Gradients in \feh\ may occur because gas leaves the
shallower potential in the dwarf galaxy's outskirts more easily than
it leaves the center.  Therefore, late star formation---from gas that
had more time to be enriched---occurs only in the dSph's center.
However, if the star formation occurs over a period shorter than the
gas redistribution timescale, then the dSph will show no
\feh\ gradient.  Our measurements of flat \feh\ radial distributions
are consistent with a short star formation duration.  Note, however,
that \citet{mar08} predicted that metallicity gradients are strongest
when stars have been forming for 1~Gyr.  After 1~Gyr, metals dispersed
by SN winds enrich the outer parts of the galaxy as much as the inner
parts.  Therefore, our interpretation of the shallow slopes is valid
only if star formation in these dSphs spanned a period significantly
shorter than 1~Gyr.  Both photometry and detailed abundances (see
\citeauthor*{kir10a}) indicate that the SF durations were indeed that
brief.

\citet{win03} has already arrived at some of these conclusions with
her Ca triplet MDFs of Sculptor, Draco, and Ursa Minor.  She detected
radial metallicity gradients in Sculptor and Draco, but not in Ursa
Minor.  She also found that the Leaky Box Model of chemical evolution
was a poor fit to the MDF of Ursa Minor, and she surmised that star
formation in Ursa Minor was particularly quick and efficient.


\section{Metallicity Trends with Luminosity}
\label{sec:trends}

So far, we have hinted at the possible role of total dwarf galaxy
luminosity in determining its chemical evolution and metallicity
distribution.  In this section, we explicitly quantify trends of the
MDF properties with luminosity.  These properties include the mean
metallicity, the intrinsic width of the MDF, and the slopes of the
radial gradients.

\begin{figure}[t!]
\centering
\includegraphics[width=\columnwidth]{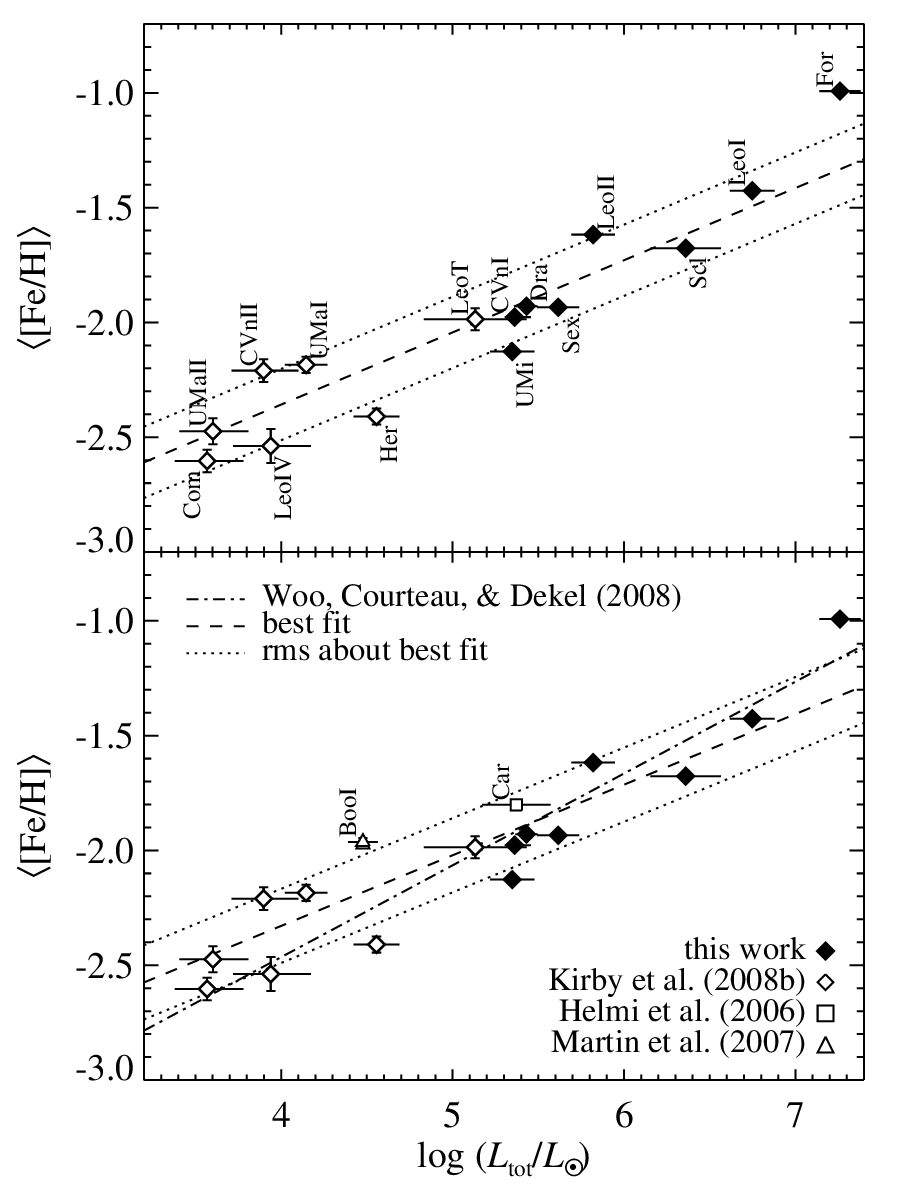}
\caption{{\it Top:} The mean [Fe/H] of MW dSphs as a function of total
  luminosity.  The dashed line is the weighted, orthogonal regression
  linear fit in $\log (L)$-[Fe/H] space, accounting for the errors in
  both $L$ and \feh\ \protect \citep{akr96}.  The dotted lines are the
  rms dispersion of the residuals.  The filled diamonds represent
  measurements from this series of papers.  The open diamonds
  represent the updated measurements of \citet{kir08b}, which were
  performed identically to those measurements presented here.  {\it
    Bottom:} Same as the top panel, with two more galaxies that were
  not measured in the same way.  Metallicity measurements for Bo{\"
    o}tes~I and Carina are based on the equivalent width of the Ca
  triplet.  The dot-dashed line is the relation of \protect
  \citet*{woo08} from Local Group galaxies, including galaxies much
  more luminous than Fornax.\label{fig:lzr}}
\end{figure}

\input{table_lzr}

Table~\ref{tab:lzr} gives a summary of the MDFs for 15 dSphs: the 8
dSphs discussed in this article and the 7 additional dSphs discussed
by \citet[][Leo~T, Hercules, Ursa Major~I and II, Leo~IV, Canes
  Venatici~II, and Coma Berenices]{kir08b}.  The MDFs of these 7 dSphs
have been updated following the changes described in Papers~I and II.
The table shows the number of stars we analyze, luminosity, and
different descriptions of the average \feh\ and width of the
\feh\ distribution.  The mean \feh\ is weighted by the inverse square
of the errors.  The MDF width $\sigma$ is given both in terms of the
standard deviation uncorrected for measurement error as well as the
width reduced by the amount of inflation caused by the estimated
measurement uncertainties.  We estimate the intrinsic spread
$\sigma(\mathfeh)$ by solving the following equation:

\begin{equation}
\frac{1}{N} \sum_{i=1}^N \frac{({\rm [Fe/H]}_i - \langle\mathfeh\rangle)^2}{(\delta{\rm [Fe/H]}_{i})^2 + \sigma(\mathfeh)^2} = 1 \: . \label{eq:fehspread}
\end{equation}

\noindent
The value of $\sigma(\mathfeh)$ for each dwarf is given in parentheses
in the column labeled $\sigma$ in Table~\ref{tab:lzr}.  The last five
columns show different shape parameters.  The median identifies the
peak of the MDF better than the mean.  The median absolute deviation
(m.a.d.) and interquartile range (IQR) are different measures of the
width of the MDF.  Skewness quantifies the asymmetry of the MDF, with
negative values indicating a metal-poor tail.  Kurtosis quantifies by
how much the MDF is peaked.  Positive kurtosis indicates that the
distribution is more sharply peaked than a Gaussian.

\subsection{Luminosity-Metallicity Relation}
\label{sec:lzr}

The average metallicity of more luminous dwarf galaxies is larger than
for fainter dwarf galaxies.  The relation between gas phase oxygen
abundances and galaxy luminosity is particularly well-studied
\citep[e.g.,][]{ski89,vad07}.  Others have studied the more basic
relation---the fundamental line---for dwarf galaxies
\citep{pra02,woo08}.  \citet{kir08b} determined the
luminosity-metallicity relation (LZR) based on medium resolution
spectral synthesis of stars in eight faint dSphs combined with Ca
triplet metallicity measurements for more luminous dSphs.  Since then,
our technique for measuring metallicities has been revised (Papers~I
and II), and we have calculated synthesis-based metallicities for the
more luminous dSphs.  The top panel of Fig.~\ref{fig:lzr} shows the
LZR for dwarf galaxies with metallicity measurements based only on
spectral synthesis.

The following equation describes the orthogonal regression fit
accounting for errors in both luminosity and \feh\ \citep{akr96},
where the errors are the standard deviations of the slope and
intercept:

\begin{equation}
\lzrkirby \: . \label{eq:lzrkirby}
\end{equation}

\noindent
The linear Pearson correlation coefficient for the data is
\pearsonkirby, indicating a highly significant correlation.

By including \feh\ measurements for other dwarfs, we may refine the
fit at the cost of losing the homogeneity of the abundance analysis.
We add Ca triplet-based measurements for Carina \citep{hel06} and
Bo{\" o}tes~I \citep{mar07}.  The bottom panel of Fig.~\ref{fig:lzr}
shows the result.  This LZR now includes all MW dwarfs less luminous
than Sagittarius except the least luminous objects (Willman~1,
\citeauthor{wil05}\ \citeyear{wil05}; Segue~1,
\citeauthor{bel07}\ \citeyear{bel07}; Segue~2,
\citeauthor{bel09}\ \citeyear{bel09}; Bo{\" o}tes~II,
\citeauthor{walsh07}\ \citeyear{walsh07}; Leo~V,
\citeauthor{bel08}\ \citeyear{bel08}; Pisces~I,
\citeauthor{wat09}\ \citeyear{wat09}; and Pisces~II and Segue~3,
\citeauthor{bel10}\ \citeyear{bel10}) because they have only a few RGB
stars, and their average metallicities are not well-determined.  For
the most part, even their luminosities are uncertain by factors of two
or more \citep{mart08a}.  As expected, the addition of two galaxies to
the existing 15 hardly changes the LZR:

\begin{equation}
\lzrall \: . \label{eq:lzrall}
\end{equation}

\noindent
The linear Pearson correlation coefficient for the data is
\pearsonall.

A straight line may be an overly simplistic model to the
luminosity-metallicity relation.  The dwarfs with $\log (L/L_\sun) >
5$ seem to lie along a steeper line than the less luminous dwarfs.  In
order to better show that difference, the dot-dashed line in
Fig.~\ref{fig:lzr} is the best-fit relation of \citet*{woo08}.  They
studied scaling relations among 41 luminous Local Group dwarf
galaxies.  They found the following relation between metal-fraction
$Z$ and stellar mass $M_*$ for dwarf ellipticals:

\begin{equation}
\log Z = -0.11 + (0.40 \pm 0.05) \log \left(\frac{M_*}{10^6 M_\sun}\right) \: . \label{eq:woo1}
\end{equation}

\noindent
Assuming, as \citeauthor{woo08}\ did, that $\mathfeh = \log
(Z/Z_\sun)$ and $Z_\sun = 0.019$ \citep{and89} and that $M_*/L =
1.36~M_\sun/L_\sun$ (the average value for their 18 dEs), we replace
Eq.~\ref{eq:woo1} with

\begin{equation}
{\rm [Fe/H]} = -2.06 + (0.40 \pm 0.05) \log \left(\frac{L_{\rm tot}}{10^5 L_\sun}\right) \: . \label{eq:woo2}
\end{equation}

\noindent
This is the dot-dashed line in Fig.~\ref{fig:lzr}.  It is an excellent
fit to the luminous half of those 17 dwarfs.  This is not surprising
because many of those dwarfs were included in \citeauthor{woo08}'s
sample.  However, the fit is not good to the dwarfs with $\log
(L/L_\sun) < 5$.

Despite the possible deviation from the LZR at low luminosities and
low metallicities, the LZR is continuous---if not linear---from
ultra-faint dSphs to massive elliptical galaxies.  \citet{tre04}
demonstrated the tight correlation between gas phase metallicity and
stellar mass over a wide range of masses.  They deduced that low mass
galaxies preferentially lose metals to galactic winds \citep{lar74b}.
Their conclusion is also consistent with the stellar
metallicity-stellar mass relation, such as derived from
spectrophotometric indices \citep[e.g.,][]{men09}.  The continuity of
the relation from massive ellipticals to galaxies with the luminosity
of Coma Berenices ($4000~L_{\sun}$) may indicate that the main
variable that dictates the amount of metals galaxies lose is galaxy
mass.

\subsection{Intrinsic [Fe/H] Spreads}

\begin{figure}[t!]
\centering
\includegraphics[width=\columnwidth]{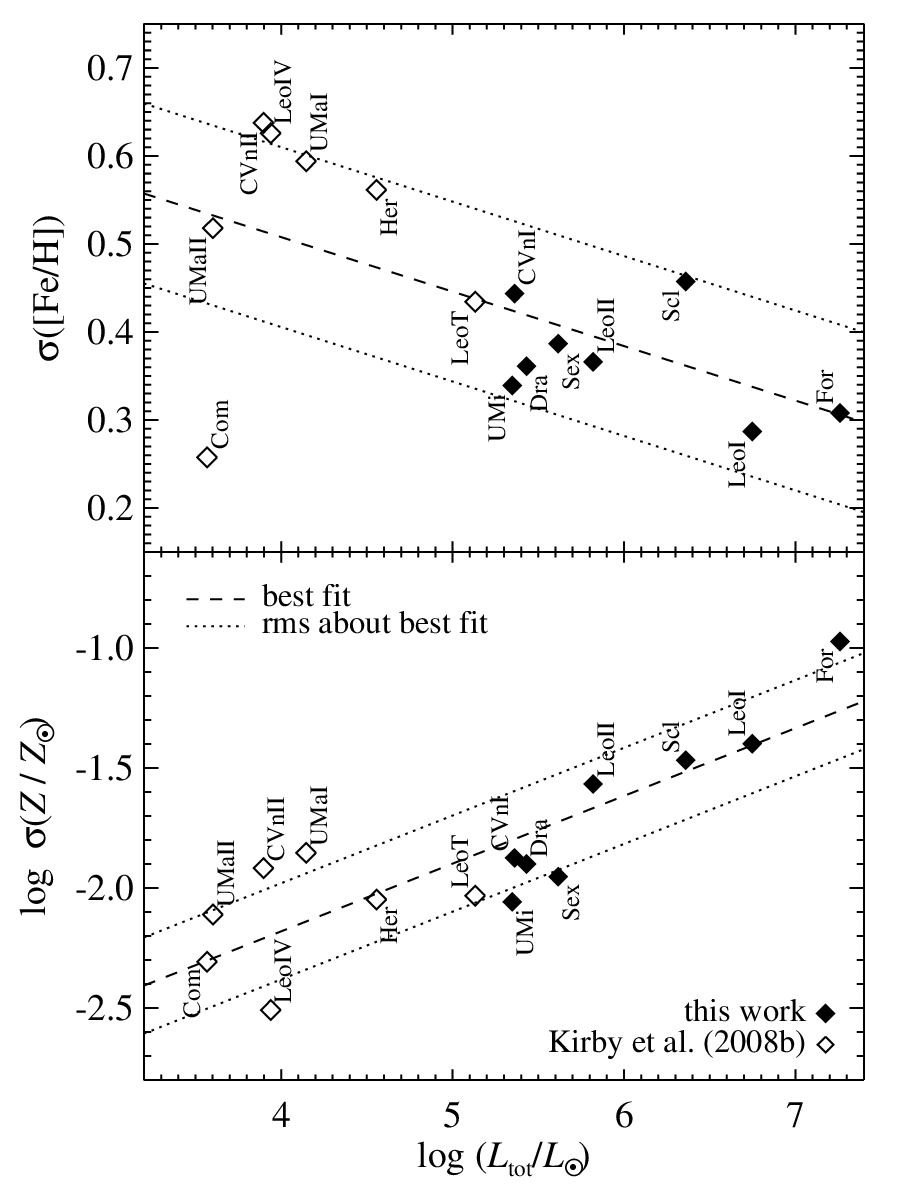}
\caption{{\it Top:} The intrinsic spread in \feh, accounting for
  measurement error, as a function of total dSph luminosity.  The
  dashed line is the least-squares fit (Eq.~\ref{eq:lzsrkirby}), and
  the dotted lines represent the rms about the dashed line.  {\it
    Bottom:} The logarithm of the linear metallicity spread as a
  function of dSph luminosity.  The dashed line is the least-squares
  fit (Eq.~\ref{eq:lzzsrkirby}).\label{fig:lzsr}}
\end{figure}

We expect the mean metallicity of a dSph to vary with luminosity
because the presence of many stars implies a history of many SNe to
enrich the gas.  The more complex enrichment histories of the more
luminous dSphs motivate us to examine how the width of the MDF varies
with luminosity.

Figure~\ref{fig:lzsr} shows the trend of $\sigma(\mathfeh)$ with dwarf
galaxy luminosity.  Intrinsic \feh\ spreads are larger in less
luminous dwarfs.  The least-squares fit is

\begin{equation}
\lzsrkirby \: . \label{eq:lzsrkirby}
\end{equation}

\noindent
The linear Pearson correlation coefficient for the data is
$\pearsonskirby$.  The coefficient is negative because
$\sigma(\mathfeh)$ is anticorrelated with luminosity.  \citet{nor10a}
also quantified the luminosity-metallicity spread-luminosity relation.
They suggested that an increase in $\sigma(\mathfeh)$ at low
luminosity indicates inhomogeneous and stochastic chemical enrichment
in the lowest mass galaxies.

However, \feh\ is a logarithmic quantity.  In order to better
visualize the physical metallicity spreads, we recast the top panel of
Fig.~\ref{fig:lzsr} in terms of the linear metal fraction: $Z/Z_{\sun}
= 10^{\mathfeh}$.  The bottom panel of Figure~\ref{fig:lzsr} shows the
logarithm of $\sigma(Z/Z_{\sun})$ versus metallicity.  The
least-squares linear fit is

\begin{equation}
\lzzsrkirby \: . \label{eq:lzzsrkirby}
\end{equation}

\noindent
The linear Pearson correlation coefficient for the data is
$\pearsonszkirby$, which indicates that the luminosity-$Z$ spread
relation is much more significant than the luminosity-\feh\ spread
relation.  

Note that $\log \sigma(Z/Z_{\sun})$ is not the same as $\sigma[\log
  (Z/Z_{\sun})] \approx \sigma(\mathfeh)$.  It is possible to
approximate $\sigma({\rm [Fe/H]})$ as $\sigma(Z)/Z$.  Instead, we
translated all values of logarithmic \feh\ into linear $Z = Z_\sun
10^{\mathfeh}$ and recalculated $\sigma(Z/Z_{\sun})$ in analogy to
Eq.~\ref{eq:fehspread}.  We estimated $\delta Z$ through standard
error propagation: $\delta Z = (Z \ln 10)\delta\mathfeh$.

The linear representation of metallicity spread shows that the
physical (linear) ranges of metallicity are somewhat similar in many
dwarfs.  Nine of the 15 dwarf galaxies cluster around $\log
\sigma(Z/Z_{\sun}) = -2$.  More interestingly, the four most luminous
dwarfs, Fornax, Leo~I, Sculptor, and Leo~II, have $\log
\sigma(Z/Z_{\sun}) > -1.6$.

These trends might be explained by the differences in star formation
duration across dwarfs.  In chemical evolution models, the width of
the MDF depends, at least in part, on the effective yield $p$.  As an
example, the width of the MDF in the Pristine Model
(Eq.~\ref{eq:pristine}) is $\sigma(Z) = \sqrt{6}p$.  For the four most
luminous dSphs, the best-fit yields are $p > 0.02$ for all three
chemical evolution models.  For the next four dSphs in order of
luminosity, the best-fit yields are $p < 0.02$ for all three models.
The effective yield $p$ encompasses gas outflow as well as SN iron
yields.  (See Table~\ref{tab:gce} for a list of the best-fit yields.)
Therefore, the luminosity-metallicity spread relation may indicate
that the more luminous dSphs more effectively retained their gas than
1the less luminous dwarf galaxies.  As a result, the more luminous
dSphs could have maintained star formation for a longer duration.  The
color-magnitude diagrams of Fornax, Leo~I, and Leo~II show that they
did indeed experience star formation more recently than 10~Gyr
\citep[e.g.,][]{mig96,buo99,sme09}, unlike the other dSphs.  Our
results from the [$\alpha$/Fe] distributions (\citeauthor*{kir10a})
also support star formation durations longer than 1~Gyr for the four
more luminous dSphs and shorter than 1~Gyr for the four less luminous
dSphs.

\subsection{Radial Gradients}

The radial gradients of the eight dSphs do not separate neatly into
more and less luminous categories.  We might reasonably guess that the
more luminous galaxies, which experienced more prolonged star
formation, would show steeper radial gradients.  However, the most
luminous dSph in our sample, Fornax, does not show a radial gradient
within the bounds of our data.  Furthermore, Sculptor and Leo~II show
very strong radial gradients even though they are near the middle of
the luminosity range of our sample of dSphs.  The explanation may
relate to Sculptor's kinematic complexity
\citep{tol04,bat08a,wal07,wal09}.  Future studies may also reveal
multiple kinematic populations in Leo~II to accompany its strong
radial gradient.

\citet{spo09} measured the radial gradients of more luminous ($M_B \le
-16.8$) galaxies in the Fornax and Virgo clusters.  The most luminous
galaxy in our sample is the Fornax dSph ($M_B = -12.6$).
\citeauthor{spo09}\ found that the magnitude of radial gradients
decreases with decreasing luminosity until $M_B \sim -17$, where the
radial gradient vanishes.  The trend with velocity dispersion is
stronger than with luminosity.  The gradients disappear by $\sim
45~\mathrm{km~s}^{-1}$, much larger than the velocity dispersion of
any galaxy in our sample.  Therefore, it is notable that half of the
galaxies in our sample (Leo~I and II, Sculptor, and Draco) display
radial gradients.  
We conclude that the radial gradients of MW dSphs do not obey the
tight relation seen for more luminous galaxies.  Instead, we speculate
that the particular star formation histories influenced by particular
interactions with the MW determine the presence or absence of radial
metallicity gradients.

At odds with \citeauthor{spo09}, \citet{kol09a} found no trend between
radial metallicity gradients and galaxy mass for galaxies in the
Fornax cluster and nearby groups.  \citet{kol09b} suggested possible
reasons for the discrepancy, but they did not find any explanations
satisfactory.  If \citeauthor{kol09a}'s result holds, then it would
not be surprising that our sample also does not show a trend between
gradients and luminosity.  Regardless, our samples are too centrally
concentrated to probe the full extent of the radial gradients for most
of the dSphs.  Other samples \citep[e.g.,][]{bat06,bat10,wal09} have
better addressed this issue.


\section{Chemical Evolution Trends with Galaxy Properties}
\label{sec:modeltrends}

\input{table_gce}

\begin{figure*}[t!]
\centering
\includegraphics[width=\textwidth]{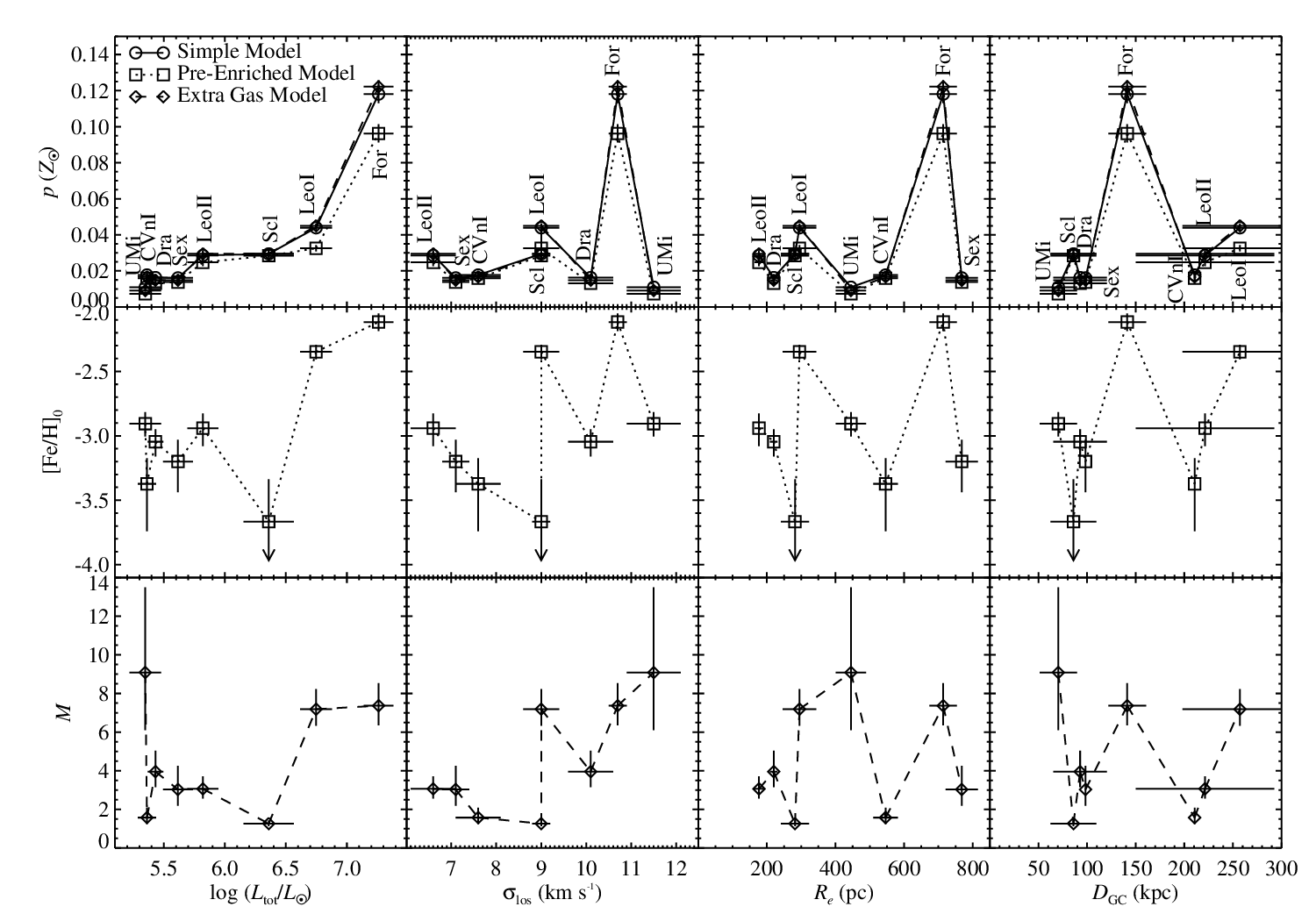}
\caption{Parameters of the best-fitting chemical evolution models
  (Table~\ref{tab:gce}) as a function of galaxy properties.  The model
  parameters are the effective metal yield ($p$) in units of the solar
  metal fraction, the initial metallicity ($\mathfeh_0$) in the
  Pre-Enriched Model, and the infall parameter ($M$) in the Extra Gas
  Model.  The galaxy properties are luminosity \citep{irw95,mart08b},
  line-of-sight velocity dispersion, two-dimensional half-light radius
  \citep[both from][and references therein]{wol10}, and Galactocentric
  distance.  The error bars represent the asymmetric 68.3\% confidence
  interval from Monte Carlo trials.  The $\mathfeh_0$ value for
  Sculptor is an upper limit.\label{fig:gce}}
\end{figure*}

The chemical evolution model fits more directly relate to the star
formation histories of the dSphs than the bulk metallicity properties.
However, the derived quantities ($p$ for the Pristine Model, $p$ and
$\mathfeh_0$ for the Pre-Enriched Model, and $p$ and $M$ for the Extra
Gas Model) are not direct observables.  In calculating them, we have
assumed that the models are good descriptions of the MDFs.  Although
we have estimated the relative goodness of fit between the Simple and
Extra Gas Models, we have not estimated the absolute goodness of fit.
Thus, the results of this section should be viewed as more directly
relevant to the star formation histories whereas the results of the
previous section should be viewed as more directly observable and
therefore more confident.

Table~\ref{tab:gce} presents the most likely chemical evolution model
parameters for each dSph along with the ratios of the maximum
likelihoods.  The two-sided uncertainties represent the 68.3\%
confidence interval (see Sec.~\ref{sec:mdf}).  Single error bars are
given for the yields because the upper error bars are equal to the
lower error bars in all cases.  Many of the Monte Carlo trials for the
Pre-Enriched Model of Sculptor reached to very low values of
$\mathrm{[Fe/H]}_0$.  The formal value for the lower error bar is
14~dex.  Therefore, we treat the calculated value of
$\mathrm{[Fe/H]}_0$ as an upper limit.

Fig.~\ref{fig:gce} shows the most likely parameters from
Table~\ref{tab:gce} plotted against the luminosity, line-of-sight
velocity dispersion, projected half-light radius, and Galactocentric
distance.  The most obvious observation is that the yield $p$
monotonically increases with dSph luminosity for the six most luminous
dSphs.  That increase is a reflection of the increasing average
metallicity and increasing metallicity spread.  For the remaining
model parameters, the dSphs may be separated into two broad
categories: more luminous, infall-dominated (Fornax, Leo~I, and
Leo~II) and less luminous, outflow-dominated (Sextans, Ursa Minor,
Draco, and Canes Venatici~I).

The more luminous, infall-dominated dSphs are more consistent with the
Extra Gas Model than the Pristine Model or the Pre-Enriched Model.  Even
so, the most likely initial metallicities for the Pre-Enriched Model
are unreasonably large.  It would be strange for Fornax and Leo~I to
have been born with a much more metal-rich ISM than other dSphs.
There is no evidence that they formed their first stars long enough
after the less luminous dSphs for the intergalactic medium to have
become so enriched.  The most likely Extra Gas Models have $M > 3$,
indicating that gas infall significantly affected star formation over
the lifetime of the galaxy.

The less luminous, outflow-dominated dSphs show similar, low effective
yields ($0.007~Z_{\sun} \le p \le 0.018~Z_{\sun}$) compared to the
more luminous dSphs, regardless of the chemical evolution model
considered.  It is possible that the average SN yields of the least
luminous dwarf galaxies are anomalous because the IMF was
stochastically sampled in tiny stellar systems \citep[e.g.,][]{koc08}.
A more likely explanation for the low values of $p$ is that gas
outflow reduced the effective yield below the value achieved by SN
ejecta.  The low masses of the less luminous dSphs rendered them
unable to retain their gas.  Gas flowed out of the galaxy from
internal mechanisms, such as SN winds, and external mechanisms, such
as ram pressure stripping.  The outflows prevented the MDFs from
achieving a high $\langle\mathfeh\rangle$ and caused the MDFs to be
more symmetric than the more luminous dSphs.

However, a curious dynamical property of dSphs may undermine this star
formation hypothesis.  The dwarf galaxies of the Local Group less
luminous than Fornax seem to inhabit dark matter halos of similar mass
\citep{mat98,pen08,str08,wal09}.  Although their stellar masses span
nearly five orders of magnitude, their dominant dark matter masses are
the same.  As a result, gas outflow from the center of, for example,
Canes Venatici~I should be no stronger than from Fornax.  However, the
masses are poorly constrained beyond the half-light radius.  It is
possible that dSph mass profiles diverge at large radii, allowing
blown-out gas to return to the more luminous galaxies with possibly
larger {\it total} mass.  However, dynamical tracers beyond 300~pc are
sparse, especially for the less luminous dSphs.  Consequently,
innovations in measuring the total mass of dSphs (such as
\citeauthor{wol10}'s \citeyear{wol10} anisotropy-independent mass
estimates or \citeauthor{amo10}'s \citeyear{amo10} phase space models)
will be necessary to better constrain the return fraction of blown-out
gas.

We have refrained from classifying Sculptor into one of these two
broad categories because none of the star formation models is a good
fit to its MDF.  It is possible that the hierarchical assembly of two
dSphs has brought together a superposition of distinct stellar
populations.  \citet{tol04} recognized this possibility in their
finding of two distinct kinematic and metallicity components in
Sculptor.  Hierarchical assembly almost certainly plays some role in
the early formation of dSphs \citep*[e.g.,][]{die07}.  However, this
process does not dominate the shapes of the MDFs for all dSphs.  For
example, Fornax and Leo~I do not show evidence of distinct metallicity
peaks.  If Fornax or Leo~I accreted so many ``sub-dwarfs'' that the
metallicity peaks are no longer distinct, then we would not expect the
average metallicities of Fornax and Leo~I to be so high and the MDF
widths to be as narrow as they are.  Instead, we conclude that
Sculptor is unique among these eight dSphs in showing the most obvious
sign of the superposition of two separate stellar populations.

With one exception, the only parameter to show a trend with any galaxy
property is the yield, which is correlated with luminosity.  The other
parameters, $\mathrm{[Fe/H]}_0$ and $M$, do not show obvious trends
with luminosity, and no parameter shows a trend with velocity
dispersion, half-light radius, or Galactocentric distance.  The
exception is that the dSphs with higher $\sigma_{\rm los}$ seem to
require larger values of $M$, or more extra gas.  It would seem that
the greater gravitational potentials of the galaxies with higher
$\sigma_{\rm los}$ attracted more external gas to power star
formation.  However, we caution against this interpretation because
the mass profiles of dSphs are complex, and $\sigma_{\rm los}$ does
not completely represent the ability of a dSph to attract additional
gas.  Instead, we regard the trend of $M$ with $\sigma_{\rm los}$ as
tenuous at best.  As an example, $\sigma_{\rm los}$ for Leo~I and
Sculptor are identical, yet they have highly discrepant values of $M$.

The chemical evolution models we have considered are overly
simplistic.  In reality, the dSphs probably have complex star
formation histories.  A steady star formation rate with instantaneous
mixing and instantaneous recycling does not completely describe the
MDF of any dSph.  Even our Extra Gas Model assumes a contrived
functional form of the gas increase (Eq.~\ref{eq:g}).  Some process,
such as interaction with the MW, must cause gas to fall into the
galaxy and trigger the star formation rate to increase one or many
times over the dSph's lifetime.  We anticipate that more realistic
semi-analytic and hydrodynamical models will provide much better
comparisons to the observed MDFs than the very simple analytic models
we have considered.


\section{Conclusions}
\label{sec:conclusions}

We have analyzed the metallicity distributions for eight dwarf
satellite galaxies of the Milky Way.  We fit analytic chemical
evolution models to the MDF of each galaxy.  A Leaky Box Model
starting from zero-metallicity gas does not faithfully describe any of
the galaxies because it encounters the same ``G dwarf problem'' that
once complicated the interpretation of the Milky Way's metallicity
distribution \citep{van62,sch63}.  A model with a fairly arbitrary
prescription for an increase in gas supply better describes the shape
of the MDFs by allowing for a narrower peak and a longer metal-poor
tail than the Leaky Box Model.  Permitting a non-zero initial
metallicity (pre-enrichment) allows the shape of the Leaky Box Model
to better fit the observed MDFs, but in no case except Canes
Venatici~I does the Pre-Enriched Model fit obviously better than the
Extra Gas Model.  In several cases, the Extra Gas Model fits much
better than the Pre-Enriched Model.

The shapes of the MDFs follow several trends with luminosity.  The
strongest trend is the luminosity-metallicity relation.  Final dwarf
galaxy luminosity can predict a dwarf galaxy's mean \feh\ to within
0.16~dex ($1\sigma$ confidence interval).  The luminosity also
determines the width of the MDF.  However, the luminosity-metallicity
spread relation is not as smooth as the luminosity-metallicity
relation.  Instead, luminosity separates the metallicity spreads into
high or low, with the four most luminous dSphs having large spreads
and dSphs with the luminosity of Sextans or smaller having small
spreads.

We surmise that dSph luminosity is a good indicator of the ability to
retain and accrete gas, despite the finding of a common central
density for all dSphs \citep{mat98,pen08,str08}.  For the more
luminous dSphs, an increase in the gas reservoir during the star
formation lifetime shapes the MDF and keeps the effective yield and
therefore the mean metallicity and metallicity spread high.  The less
luminous dSphs are less able to retain gas that leaves via supernova
winds or interaction with the Milky Way.  Finally, all of the chemical
evolution models we consider are much too narrow to explain the MDF
for Sculptor.  The previous evidence for multiple kinematic
populations in Sculptor \citep{tol04,bat08a} suggests that Sculptor
experienced at least two distinct, major episodes of star formation.
We have not considered kinematics in our analysis, but we recognize
the value radial velocities add to abundance data.  We plan to explore
the relationship between velocity dispersion, \feh, and [$\alpha$/Fe]
in the future.

Stellar mass, not luminosity, is likely the independent variable, a
concept that \citet{woo08} explored in detail.  The conversion from
luminosity to stellar mass involves the ages and metallicities of the
component populations.  Rather than complicate the observational data,
we have chosen to present the dSph metallicity relations against
luminosity because it is a direct observable.  Most of the dSphs in
our sample are ancient and metal-poor.  As a result, the relation
between their luminosities and stellar masses is one-to-one.  Three
dSphs---Fornax and Leo~I and II---have younger populations, which
result in higher luminosities at a given stellar mass.  Nonetheless,
the relation between luminosity and stellar mass is roughly monotonic
for the dSphs in our sample, according to the stellar masses derived
by \citet{woo08} or \citet{orb08}.  Therefore, we conclude that
luminosity is a good proxy for the more fundamental parameter, stellar
mass, for these eight dSphs.

The radial gradients of \feh\ do not obey a relation with total dSph
luminosity.  The slopes of the gradients for four dSphs are consistent
with zero, and four more are significantly negative.  The slopes for
Sculptor and Leo~II are at least as steep as $-0.18$~dex per core
radius.  Negative slopes are to be expected for most galaxies.
Consequently, we have no satisfying explanation for the lack of a
pattern for which dSphs happen to show radial gradients, even though
we have characterized the star formation of the four less luminous
dSphs as dominated by gas outflow.  Samples more radially extended
than ours might show metallicity gradients in all dSphs.

Although we have assigned quantitative parameters for analytic
chemical evolution models to each galaxy, we do not believe that any
of the chemical evolution models are excellent fits to any dSph.  In
particular, the assumptions of instantaneous mixing and recycling are
inappropriate for small galaxies with star formation lifetimes longer
than the timescale for Type~Ia supernova explosions.  We expect that
models that incorporate inhomogeneous pockets of star formation
\citep[such as the models of][]{mar06,mar08,rev09} and time-delayed
iron enhancement from Type~Ia SNe \citep[such as the models
  of][]{lan04,mar06,mar08,rev09} will much better describe the dSph
MDFs.  Interested modelers wishing to compare their predictions to our
observed MDFs may find the complete catalog of \feh\ measurements in
\citeauthor*{kir10b}.

\acknowledgments We thank the anonymous referee for helpful advice
that improved this manuscript.  We also thank Bob Kraft for helpful
comments and Julianne Dalcanton for the suggestion that gas may become
available for star formation in ways other than the infall of external
gas.  We also recognize the work of Marla Geha, Steve Majewski, Connie
Rockosi, Michael Siegel, Chris Sneden, Tony Sohn, and Peter Stetson in
making the data catalog used in this article possible.  Support for
this work was provided by NASA through Hubble Fellowship grant
HST-HF-51256.01 awarded to ENK by the Space Telescope Science
Institute, which is operated by the Association of Universities for
Research in Astronomy, Inc., for NASA, under contract NAS 5-26555.
GAL acknowledges financial support from the Brazilian agency FAPESP
(proj.\ 06/57824-1).  NSF grant AST-0908139, awarded to JGC, provided
partial support for this project.  PG acknowledges NSF grants
AST-0307966, AST-0607852, and AST-0507483.

The authors wish to recognize and acknowledge the very significant
cultural role and reverence that the summit of Mauna Kea has always
had within the indigenous Hawaiian community.  We are most fortunate
to have the opportunity to conduct observations from this mountain.

{\it Facility:} \facility{Keck:II (DEIMOS)}

\end{document}

%% file: table_gradients.tex
\setcounter{table}{0}
\begin{deluxetable*}{lcccccc}
\tabletypesize{\small}
\tablecolumns{7}
\tablewidth{0pt}
\tablecaption{Metallicity Gradients\label{tab:gradients}}
\tablehead{\colhead{dSph} & \colhead{Distance} & \colhead{$r_c$} & \colhead{$r_c$} & \colhead{$d{\rm [Fe/H]} / d\theta$} & \colhead{$d{\rm [Fe/H]} / dr$} & \colhead{$d{\rm [Fe/H]} / d(r/r_c)$} \\
 & (kpc) & (arcmin) & (pc) & (dex deg$^{-1}$) & (dex kpc$^{-1}$) & (dex)}
\startdata
Fornax            &             139 &            13.7 &             550 & $-0.07 \pm 0.09$ & $-0.03 \pm 0.04$ & $-0.02 \pm 0.02$ \\
Leo~I             &             254 &        \phn 3.3 &             240 & $-2.01 \pm 0.10$ & $-0.45 \pm 0.02$ & $-0.11 \pm 0.01$ \\
Sculptor          &         \phn 85 &        \phn 5.8 &             140 & $-1.86 \pm 0.16$ & $-1.24 \pm 0.11$ & $-0.18 \pm 0.02$ \\
Leo~II            &             219 &        \phn 2.9 &             180 & $-4.26 \pm 0.31$ & $-1.11 \pm 0.08$ & $-0.21 \pm 0.01$ \\
Sextans           &         \phn 95 &            16.6 &             460 & $+0.20 \pm 0.19$ & $+0.12 \pm 0.11$ & $+0.06 \pm 0.05$ \\
Draco             &         \phn 92 &        \phn 9.0 &             240 & $-0.73 \pm 0.13$ & $-0.45 \pm 0.08$ & $-0.11 \pm 0.02$ \\
Canes Venatici~I  &             210 &        \phn 8.9\tablenotemark{a}           &             540\tablenotemark{a}           & $-0.48 \pm 0.36$ & $-0.13 \pm 0.10$ & $-0.07 \pm 0.05$ \\
Ursa Minor        &         \phn 69 &            15.8 &             320 & $-0.21 \pm 0.19$ & $-0.18 \pm 0.16$ & $-0.06 \pm 0.05$ \\
\enddata
\tablerefs{Distances adopted from \citet{riz07} for Fornax, \citet{bel04} for Leo~I, \citet{pie08} for Sculptor, \citet{sie10} for Leo~II, \citet{lee03} for Sextans, \citet{bel02} for Draco, \citet{kue08} for Canes Venatici~I, and \citet{mig99} for Ursa Minor.  Core radii adopted from \citet{irw95} except for Canes Venatici I, for which we adopt the half-light radius derived by \citet{mart08a}.}
\tablenotetext{a}{Half-light radius instead of core radius.}
\end{deluxetable*}

%% file: table_lzr.tex
\begin{deluxetable*}{lccccccccc}
\tablecolumns{10}
\tablewidth{0pt}
\tablecaption{Summary of dSph MDFs\label{tab:lzr}}
\tablehead{\colhead{dSph} & \colhead{\phantom{\tablenotemark{a}}$N$\tablenotemark{a}} & \colhead{$\log (L/L_{\sun})$} & \colhead{\phantom{\tablenotemark{b}}$\langle {\rm [Fe/H]}\rangle$\tablenotemark{b}} & \colhead{\phantom{\tablenotemark{c}}$\sigma$\tablenotemark{c}} & \colhead{Median} & \colhead{\phantom{\tablenotemark{d}}m.a.d.\tablenotemark{d}} & \colhead{\phantom{\tablenotemark{e}}IQR\tablenotemark{e}} & \colhead{Skewness} & \colhead{\phantom{\tablenotemark{f}}Kurtosis\tablenotemark{f}}}
\startdata
Fornax            &         675 & $7.3 \pm 0.1$ & $-0.99 \pm 0.01$ & $0.36$ $(0.31)$ & $-1.01$ & $0.19$ & $0.37$ &     $-1.33 \pm 0.09$ & \phs$ 3.58 \pm 0.19$ \\
Leo~I             &         827 & $6.7 \pm 0.1$ & $-1.43 \pm 0.01$ & $0.33$ $(0.29)$ & $-1.42$ & $0.18$ & $0.37$ &     $-1.47 \pm 0.09$ & \phs$ 4.99 \pm 0.17$ \\
Sculptor          &         376 & $6.4 \pm 0.2$ & $-1.68 \pm 0.01$ & $0.48$ $(0.46)$ & $-1.67$ & $0.37$ & $0.75$ &     $-0.67 \pm 0.13$ & \phs$ 0.25 \pm 0.25$ \\
Leo~II            &         258 & $5.8 \pm 0.1$ & $-1.62 \pm 0.01$ & $0.42$ $(0.37)$ & $-1.59$ & $0.23$ & $0.51$ &     $-1.11 \pm 0.15$ & \phs$ 1.10 \pm 0.30$ \\
Sextans           &         141 & $5.6 \pm 0.1$ & $-1.93 \pm 0.01$ & $0.48$ $(0.39)$ & $-2.00$ & $0.29$ & $0.57$ &     $-0.10 \pm 0.20$ & \phs$ 0.47 \pm 0.41$ \\
Draco             &         298 & $5.4 \pm 0.1$ & $-1.93 \pm 0.01$ & $0.47$ $(0.36)$ & $-1.93$ & $0.26$ & $0.51$ &     $-0.51 \pm 0.14$ & \phs$ 0.73 \pm 0.28$ \\
Canes Venatici~I  &         174 & $5.4 \pm 0.1$ & $-1.98 \pm 0.01$ & $0.55$ $(0.44)$ & $-1.98$ & $0.36$ & $0.71$ &     $-0.26 \pm 0.18$ & \phs$ 0.23 \pm 0.37$ \\
Ursa Minor        &         212 & $5.3 \pm 0.1$ & $-2.13 \pm 0.01$ & $0.47$ $(0.34)$ & $-2.13$ & $0.25$ & $0.50$ &     $-0.03 \pm 0.17$ & \phs$ 1.34 \pm 0.33$ \\
Leo~T             &     \phn 18 & $5.1 \pm 0.3$ & $-1.99 \pm 0.05$ & $0.52$ $(0.43)$ & $-1.92$ & $0.31$ & $0.79$ &     $-0.53 \pm 0.54$ &     $-1.38 \pm 1.04$ \\
Hercules          &     \phn 21 & $4.6 \pm 0.1$ & $-2.41 \pm 0.04$ & $0.64$ $(0.56)$ & $-2.62$ & $0.46$ & $0.96$ & \phs$ 0.51 \pm 0.50$ &     $-0.94 \pm 0.97$ \\
Ursa Major~I      &     \phn 28 & $4.1 \pm 0.1$ & $-2.18 \pm 0.04$ & $0.64$ $(0.59)$ & $-2.62$ & $0.30$ & $0.60$ & \phs$ 0.61 \pm 0.44$ &     $-0.62 \pm 0.86$ \\
Leo~IV            &     \phn 12 & $3.9 \pm 0.2$ & $-2.54 \pm 0.07$ & $0.70$ $(0.63)$ & $-2.35$ & $0.58$ & $0.86$ & \phs$ 0.61 \pm 0.64$ &     $-0.74 \pm 1.23$ \\
Canes Venatici~II &     \phn 15 & $3.9 \pm 0.2$ & $-2.21 \pm 0.05$ & $0.71$ $(0.64)$ & $-2.68$ & $0.34$ & $0.62$ & \phs$ 0.43 \pm 0.58$ &     $-0.37 \pm 1.12$ \\
Ursa Major~II     & \phn\phn  9 & $3.6 \pm 0.2$ & $-2.47 \pm 0.06$ & $0.57$ $(0.52)$ & $-2.41$ & $0.41$ & $0.92$ & \phs$ 0.62 \pm 0.72$ &     $-0.69 \pm 1.40$ \\
Coma Berenices    &     \phn 18 & $3.6 \pm 0.2$ & $-2.60 \pm 0.05$ & $0.40$ $(0.26)$ & $-2.70$ & $0.29$ & $0.56$ &     $-0.20 \pm 0.54$ &     $-1.02 \pm 1.04$ \\
\enddata
\tablerefs{To derive the luminosities of Fornax through Draco and Ursa Minor, we adopt the integrated $V$-band magnitudes of \citet{irw95} and the distances given in Table~\ref{tab:gradients}.  For the other galaxies, we adopt the luminosities of \citet{mart08a}.}
\tablenotetext{a}{Number of member stars, confirmed by radial velocity, with measured [Fe/H].}
\tablenotetext{b}{Mean [Fe/H] weighted by the inverse square of estimated measurement uncertainties.}
\tablenotetext{c}{The number in parentheses is the intrinsic \feh\ spread corrected for measurement uncertainties.}
\tablenotetext{d}{Median absolute deviation.}
\tablenotetext{e}{Interquartile range.}
\tablenotetext{f}{Actually the excess kurtosis, or 3 less than the raw kurtosis.  This quantifies the degree to which the distribution is more sharply peaked than a Gaussian.}
\end{deluxetable*}

%% file: table_gce.tex
\begin{deluxetable*}{lccccccccc}
\tabletypesize{\footnotesize}
\tablecolumns{10}
\tablewidth{0pt}
\tablecaption{Chemical Evolution Models\label{tab:gce}}
\tablehead{ & Pristine Model & & \multicolumn{2}{c}{Pre-Enriched Model} & & \multicolumn{2}{c}{Extra Gas Model} & & \\ \cline{2-2} \cline{4-5} \cline{7-8}
\colhead{dSph} & \colhead{$p$ ($Z_\sun$)} & & \colhead{$p$ ($Z_\sun$)}
& \colhead{${\rm [Fe/H]}_0$} & & \colhead{$p$ ($Z_\sun$)} & \colhead{$M$} & \colhead{$\ln \frac{L_{\rm max}({\rm Pre\mbox{-}Enriched})}{L_{\rm max}({\rm Pristine})}$} & \colhead{$\ln \frac{L_{\rm max}({\rm Extra~Gas})}{L_{\rm max}({\rm Pristine})}$}}
\startdata
Fornax        & $0.118 \pm 0.005$         & & $0.096 \pm 0.005$         & \phs $-2.12 \pm 0.07$     & & $0.122 \pm 0.004$         &    $ 7.4_{-1.0}^{+1.2}$ &     \phs\phn $ 49.98$ &          $125.62$ \\
Leo I         & $0.044 \pm 0.002$         & & $0.033 \pm 0.002$         & \phs $-2.35_{-0.06}^{+0.05}$ & & $0.045 \pm 0.001$         &    $ 7.2_{-0.9}^{+1.0}$ &     \phs\phn $ 81.45$ &          $163.32$ \\
Sculptor      & $0.029 \pm 0.002$         & & $0.028 \pm 0.002$         & $< -3.67^{+0.33}$          & & $0.029 \pm 0.002$         &    $ 1.3_{-0.1}^{+0.2}$ &     \phn\phn $ -0.11$ & \phn\phn $  1.59$ \\
Leo II        & $0.029 \pm 0.002$         & & $0.025 \pm 0.002$         & \phs $-2.94_{-0.14}^{+0.11}$ & & $0.030 \pm 0.002$         &    $ 3.1_{-0.5}^{+0.6}$ &     \phs\phn $ 12.67$ &     \phn $ 21.08$ \\
Sextans       & $0.016 \pm 0.002$         & & $0.014 \pm 0.002$         & \phs $-3.20_{-0.24}^{+0.17}$ & & $0.015 \pm 0.001$         &    $ 3.0_{-0.8}^{+1.2}$ & \phs\phn\phn $  6.43$ & \phn\phn $  5.62$ \\
Draco         & $0.016 \pm 0.001$         & & $0.013 \pm 0.001$         & \phs $-3.05_{-0.11}^{+0.10}$ & & $0.015 \pm 0.001$         &    $ 4.0_{-0.8}^{+1.1}$ &     \phs\phn $ 17.51$ &     \phn $ 20.34$ \\
Can.\ Ven.\ I & $0.018 \pm 0.002$         & & $0.016 \pm 0.002$         & \phs $-3.37_{-0.37}^{+0.20}$ & & $0.017 \pm 0.002$         &    $ 1.6_{-0.3}^{+0.5}$ & \phs\phn\phn $  4.55$ & \phn\phn $  0.26$ \\
Ursa Minor    & $0.011 \pm 0.001$         & & $0.007 \pm 0.001$         & \phs $-2.91_{-0.10}^{+0.09}$ & & $0.009 \pm 0.001$         &    $ 9.1_{-3.0}^{+4.4}$ &     \phs\phn $ 22.10$ &     \phn $ 21.63$ \\
\enddata
\end{deluxetable*}